\documentclass{article}
\usepackage{jheppub}
\usepackage{graphicx} 
\usepackage{graphbox}
\usepackage[usenames,dvipsnames]{xcolor} 
\usepackage{tikz}	
\usepackage{dirtytalk}
\usepackage{dsfont}
\usepackage{amsthm}
\usepackage{float}
\usepackage{bbm, dsfont}
\usepackage[normalem]{ulem}
\usepackage{slashed}
\usepackage{mathtools}
\usepackage{cancel}
\usepackage{cleveref}
\usepackage{orcidlink}
\usepackage{color}
\usepackage{colortbl}
\usepackage{tcolorbox}
\usepackage{enumerate}
\usepackage{fontawesome}
\usepackage{adjustbox}
\usepackage{tcolorbox}
\usepackage[T1]{fontenc}
\usepackage{newtxtext,newtxmath}
\usepackage{cancel}
\usepackage{nicematrix}

\usepackage{bm}

\title{On an approach to canonicalizing elliptic Feynman integrals}

\abstract{We present generic expressions for the integrands of canonical bases under maximal cut in elliptic Feynman integral families with multiple kinematic scales. Such integrals frequently arise in phenomenologically relevant scattering processes. The derivation of our results starts from the Legendre normal form of elliptic curves, where the geometric properties of the curves are simple and explicit, and further kinematic singularities are presented as marked points. The simplicity of the normal form allows a straightforward construction of canonical bases with an arbitrary number of marked points. They can then be mapped into any univariate elliptic integral families via an appropriate M\"obius transformation, leading to universal expressions for the integrands. As a demonstration, we discuss the application of our method to several concrete examples, including two new integral families whose canonical bases were not available in the literature. In several examples, we derive canonical bases for the full integral families without any cuts, demonstrating the simplicity of the sub-sector dependence of our canonical bases.}

\author[b,c,\orcidlink{0000-0003-3537-3846}]{Jiaqi Chen,}
\emailAdd{jiaqichen@cup.edu.cn}
\author[a,\orcidlink{0000-0001-7707-8138}]{Li Lin Yang,}
\emailAdd{yanglilin@zju.edu.cn}
\author[a,\orcidlink{0009-0005-3972-2611}]{Yiyang Zhang}
\emailAdd{yiyangzhang@zju.edu.cn}
\affiliation[a]{
    Zhejiang Institute of Modern Physics, 
    School of Physics, Zhejiang University, \\
    Hangzhou 310058, China
    }
\affiliation[b]{
    Beijing Key Laboratory of Optical Detection Technology for Oil and Gas, China University of Petroleum-Beijing, \\
    Beijing 102249, China
    }
\affiliation[c]{
    Basic Research Center for Energy Interdisciplinary, College of Science, China University of Petroleum-Beijing, \\
    Beijing 102249, China
    }

\allowdisplaybreaks
\begin{document}
% \preprint{}

\maketitle
\normalem
\allowdisplaybreaks
\raggedbottom

\newpage

\section{Introduction}

\label{sec_intro}

With the construction of the High Luminosity-Large Hadron Collider (HL-LHC)~\cite{Cepeda:2019klc} and prospective high-energy colliders such as the Circular Electron Positron Collider (CEPC)~\cite{CEPCStudyGroup:2018ghi} and the Future Circular Collider (FCC)~\cite{FCC:2018byv,FCC:2018evy,FCC:2018vvp}, which promise greater luminosity and energy, we are on the verge of exploring fundamental natural phenomena with unprecedented precision. These experimental advancements provide a robust testing ground for the Standard Model (SM), allowing physicists to rigorously validate its predictions, identify potential limitations, and perhaps even discover signs of new physics.

To fully exploit the anticipated wealth of experimental data, highly accurate theoretical predictions are essential. Such predictions are grounded in quantum field theory (QFT) and are typically obtained through perturbative expansion, which necessitates calculating a variety of Feynman integrals. Understanding the mathematical structure of these integrals is crucial to their computation (see~\cite{Weinzierl:2022eaz,Abreu:2022mfk,Badger:2023eqz} for recent reviews). Each integral family spans a finite-dimensional linear space, with its basis integrals known as \textit{master integrals} (MIs). Any integral within the family can be expressed as a linear combination of MIs, often through \textit{integration-by-parts} (IBP) identities~\cite{Tkachov:1981wb,Chetyrkin:1981qh}, which are systematically solved using the Laporta algorithm~\cite{Laporta:2000dsw}. Several public software packages, such as \texttt{Reduze}~\cite{Studerus:2009ye,vonManteuffel:2012np}, \texttt{LiteRed}~\cite{Lee:2012cn,Lee:2013mka}, \texttt{FIRE}~\cite{Smirnov:2019qkx,Smirnov:2023yhb}, and \texttt{Kira}~\cite{Maierhofer:2017gsa,Klappert:2020nbg}, implement this algorithm. There are also efforts to work in a smaller linear equations system, such as \texttt{NeatIBP}~\cite{Wu:2023upw,Wu:2025aeg} based on \textit{syzygy} method and \texttt{Blade}~\cite{Guan:2024byi} based on the method of \textit{block-triangular form}. Recently, Feynman integrals have been reinterpreted within \textit{twisted (co-)homology} as \textit{twisted periods}~\cite{Aomoto:2011ggg}, enabling reduction via \textit{intersection numbers}~\cite{Mastrolia:2018uzb,Frellesvig:2019uqt} rather than traditional IBP equations.

The \textit{differential equations} method~\cite{Kotikov:1990kg,Kotikov:1991pm,Remiddi:1997ny,Gehrmann:1999as}, and specifically the \textit{canonical differential equations} approach~\cite{Henn:2013pwa}, represents the state-of-the-art method for calculating MIs analytically. This approach leverages an \textit{$\varepsilon$-factorized form}, allowing MIs to be expressed as Laurent expansions in the dimensional regulator $\varepsilon$, with Chen’s iterated integrals~\cite{Chen:1977oja} as coefficients.\footnote{A canonical basis has other nice properties in addition to $\varepsilon$-factorization.} The simplest iterated integrals found in Feynman integrals are \textit{multiple polylogarithms} (MPLs)~\cite{Goncharov:1998kja,Goncharov:2001iea,Vollinga:2004sn}, which are iterated integrals over the genus-$0$ Riemann sphere with marked points, or say over \textit{moduli space} ${\cal M}_{0,n}$~\cite{Bogner:2014mha}. For MPL integral families, canonical differential equations can be achieved through rational and algebraic transformations alone, and several tools are available to assist in identifying such transformations~\cite{Gituliar:2017vzm,Prausa:2017ltv,Meyer:2017joq,Lee:2020zfb,Dlapa:2020cwj}. An alternative approach constructs canonical bases through \textit{${\rm d} \log$-forms}~\cite{Henn:2020lye,Chen:2020uyk,Chen:2022lzr}.\footnote{However, not all integrals in ${\rm d} \log$-forms are polylogarithmic, see~\cite{Duhr:2020gdd}.} Basis integrals derived this way automatically possess \textit{uniform transcendentality} (UT)~\cite{Henn:2013pwa}, which is another key property of canonical bases in addition to $\varepsilon$-factorization. MPLs exhibit properties that facilitate algorithmic extraction of critical information, aiding in the construction of canonical bases and enriching our understanding of their mathematical structure.

However, as the complexity of the problem increases with additional loops and mass scales, the MPL functions space becomes insufficient to accommodate all integrals within an integral family (see, e.g.,~\cite{Bourjaily:2022bwx} for a recent review). In these cases, elliptic integrals—the next class of iterated integrals—emerge, first noted in the context of the two-loop electron self-energy in QED~\cite{Sabry:1962rge}. Fully analytic results for this case were obtained only half a century later~\cite{Honemann:2018mrb}, and additional significant calculations involving more scales or higher loops are found in works such as~\cite{Bogner:2019lfa,Giroux:2024yxu,Pogel:2022yat,Pogel:2022ken,Pogel:2022vat}. Three-loop corrections have also been computed in recent studies~\cite{Duhr:2024bzt}. Elliptic Feynman integrals are iterated integrals over a genus-$1$ Riemann surface, or say torus, with marked points, equivalently, over moduli space ${\cal M}_{1,n}$~\cite{Weinzierl:2020kyq}. Unlike MPLs, which are defined on the Riemann sphere with a unique shape, the genus-$1$ Riemann surface underlying elliptic Feynman integrals varies and can be parameterized by a modular variable, $\tau$. To express the $\varepsilon$-expansion of elliptic Feynman integrals, we need both the \textit{elliptic multiple polylogarithms} (eMPLs)~\cite{Levin:2007tto,Brown:2011wfj,Broedel:2017kkb} and iterated integrals of \textit{modular forms} or \textit{Eisenstein series}~\cite{Adams:2017ejb,Broedel:2018iwv,Duhr:2019rrs,Weinzierl:2020kyq}. It's well-known that the underlying geometry for an integral can be made explicit with the analysis of \textit{leading singularities}~\cite{Cachazo:2008vp,Arkani-Hamed:2010pyv}. Specifically, the underlying elliptic curve of an elliptic sector can be identified in \textit{(loop-by-loop) Baikov representations}~\cite{Baikov:1996rk,Baikov:1996iu,Frellesvig:2017aai,Harley:2017qut} through the \textit{maximal cut}~\cite{Frellesvig:2017aai,Harley:2017qut,Primo:2016ebd,Bosma:2017ens,Primo:2017ipr}. Implementations for (loop-by-loop) Baikov representations are offered by~\cite{Frellesvig:2017aai,Jiang:2023qnl,Jiang:2024eaj,Frellesvig:2024ymq,Correia:2025yao}. Alternatively, we can study the elliptic curve by analyzing the \textit{periods} of the elliptic curve through solutions to a second-order irreducible \textit{Picard-Fuchs operator}~\cite{Adams:2017tga,Adams:2018bsn,Adams:2018kez}. 

Due to the success of the method of canonical differential equations in solving the MPL integral families, there have been substantial efforts to extend the method to the elliptic cases appearing in various scattering processes \cite{Muller:2022gec,Jiang:2023jmk,Gorges:2023zgv,Delto:2023kqv,Wang:2024ilc,Forner:2024ojj,Schwanemann:2024kbg,Marzucca:2025eak,Becchetti:2025rrz,Becchetti:2025oyb}. In such cases, there are subtleties in the precise definition of ``canonical'' \cite{Broedel:2018iwv,Broedel:2018qkq,Broedel:2019hyg,Dlapa:2022wdu,Frellesvig:2023iwr,Duhr:2024uid,Duhr:2025lbz}, but one may at least attempt to obtain an $\varepsilon$-factorized system of differential equations which still offers simplification in the calculations. There are now several methods for deriving $\varepsilon$-factorized bases for elliptic cases~\cite{vonManteuffel:2017hms,Adams:2018bsn,Frellesvig:2021hkr,Pogel:2022vat,Giroux:2022wav,Dlapa:2022wdu,Gorges:2023zgv}, and some of them can even be extended to more complex geometries, such as \textit{Calabi-Yau varieties}. In the elliptic cases, these methods usually construct pre-canonical bases according to the elliptic curves defined by some degree-3 or degree-4 polynomials, where other kinematic singularities appear as marked points. Further transformations are then applied to convert the differential equations for these pre-canonical bases to the $\varepsilon$-factorized form.

In this paper, we present universal expressions for the integrands of canonical bases in any univariate elliptic integral families under the maximal cut. The derivation of the integrands utilizes the \textit{Legendre normal form}\footnote{The Legendre normal form has also been employed in \cite{Broedel:2018rwm}. Similarly, the \textit{Rosenhain normal form} for the hyperelliptic case has been explored recently in \cite{Duhr:2024uid}.} of elliptic curves, where the geometric information is evident through the elliptic modular lambda function $\lambda$. The results are then mapped to a generic elliptic family via a M\"obius transformation. We use the term ``canonical'' here since they exhibit many of the important properties of canonical bases advocated by the literature~\cite{Dlapa:2022wdu,Frellesvig:2023iwr,Duhr:2024uid}, besides $\varepsilon$-factorization. In particular, they approach $\mathrm{d} \log$ forms at the degenerate limit of the elliptic curves, where the differential equations also exhibit a simple pole behavior. 

Our results are at the integrand level, and do not require further information from the differential equations within the top sector. We apply our method to several concrete examples, including the unequal-mass sunrise family and non-planar double-box families. We further consider two phenomenologically relevant integral families, and provide their canonical bases for the first time. In several examples, we obtain canonical bases for the full integral families without any cuts. This benefits from the simplicity of the sub-sector dependence of our pre-canonical bases. To further simplify the sub-sector dependence, we make preliminary attempts for multivariate constructions under the next-to-maximal cut, where we show that a large portion of the differential equations can be automatically rendered $\varepsilon$-factorized.

This paper is organized as follows. In Sec.~\ref{sec_construction}, we present the main results,  the universal expressions for the canonical bases, and their derivation. We apply the results to several examples in Sec.~\ref{sec_eg}. The summary and outlook are provided in Sec.~\ref{sec_conclusion}. Several appendices are included at the end of the paper. A different strategy \cite{vonManteuffel:2017hms,Frellesvig:2021hkr} for deriving an $\varepsilon$-factorized basis is presented in App.~\ref{app_period_eps}. The construction using the Jacobi normal form of elliptic curves is discussed in App.~\ref{app_jacobi}. Some lengthy expressions relevant to the example in Sec.~\ref{subsec_npdb} are collected in App.~\ref{app_basis}. Special relations between elliptic integrals are compiled in App.~\ref{app_special}. The canonical bases for all examples are offered in the attached ancillary files.

\section{Canonical basis under the maximal cut}

\label{sec_construction}

\subsection{Integral family and canonical basis}
\label{sec:canbasis}

An $L$-loop Feynman integral family with $E + 1$ external legs is defined as
\begin{equation} 
  \label{FIs}
  I_{\nu_1\cdots\nu_N} 
  = e^{L\varepsilon\gamma_E} (\mu^2)^{\nu - L d / 2} \int \, \bigg[\,\prod_{j = 1}^L \frac{{\rm d} ^d k_j}{i \pi^{d/2}}\,\bigg] \frac{1}{D_1^{\nu_1}\cdots D_N^{\nu_N}} \, ,
\end{equation}
with $N=L(L+1)/2+LE$, $d=d_0-2\varepsilon$ ($d_0\in2\mathbb{Z}^{+}$) and $\nu=\sum_{i = 1}^N \nu_i$. A sector in the family is specified by the set of $D_i$ with the corresponding $\nu_i$ positive, and the rest of the propagators are \textit{irreducible scalar products} (ISPs). Unless otherwise stated, we will consider integral families in $d = 4 - 2 \varepsilon$.

We consider an elliptic sector within this integral family in the (loop-by-loop) Baikov representation \cite{Frellesvig:2017aai,Jiang:2023qnl,Jiang:2024eaj,Frellesvig:2024ymq,Correia:2025yao}. We assume that there is only one ISP variable $z$ in this representation, such that after applying maximal cut, the integrals in this sector can be expressed as
\begin{equation}
  I 
  = \int_{\cal C} u \, \varphi
  \, , 
\end{equation}
where the multi-valued function $u$ (the twist in the language of twisted cohomology) is\footnote{Here and in the following, we always suppress factors that only depend on kinematic variables in $u$.}
\begin{equation}
  \label{eq:general_u}
  u = u(z) = [P_4 (z)]^{- 1 / 2} \prod_{i = 1}^n (z - c_i)^{- \beta_i \varepsilon} 
  \, , 
  \quad
  P_4(z) \equiv (z-c_1)(z-c_2)(z-c_3)(z-c_4) \,,
\end{equation}
where the branch points $c_i$'s are distinct and the corresponding $\beta_i$'s are integers. The single-valued differential $1$-form $\varphi$ may have poles at some of the branch points $c_i$'s. Note that there is an implicit branch point at infinity, which will be denoted as $c_{n + 1}  \equiv c_{\infty} \equiv \infty$. For convenience, we assume $c_1 < c_2 < c_3 < c_4$, and introduce the reduced twist
\begin{equation}
  \bar{u}(z) = \prod_{i = 1}^n (z - c_i)^{- \beta_i \varepsilon} \,,
\end{equation}
so that the integrals can be represented as
\begin{equation}
  \label{eq:integral_family}
  I = \int_{\cal C} \bar{u}(z) \frac{\varphi}{\sqrt{P_4(z)}} \equiv \int_{\cal C} \bar{u}(z) \, \phi \,.
\end{equation}

The degree-4 polynomial $P_4(z)$ defines an elliptic curve $w^2 = P_4(z)$. The shape of the elliptic curve is characterized by the modular lambda $\lambda$,
\begin{equation}
  \label{lambda}
  \lambda = \frac{c_{14} c_{23}}{c_{13} c_{24}} \, , 
  \quad (c_{ij} \equiv c_i - c_j) \, .
\end{equation}
The elliptic curve can be transformed into the Legendre normal form
\begin{equation}
  \label{legendre}
  y^2  = P_L(x) \equiv
  x  (x - \lambda) (x - 1) \, ,
\end{equation}
by a M\"obius transformation
\begin{equation}
  \label{M\"obius}
  z
  \mapsto x 
  = T (z) 
  = \frac{(z - c_2) c_{14}}{(z - c_1) c_{24}}
  \, .
\end{equation}
Under this M\"obius transformation, the branch points are mapped as
\begin{align}
  \label{z2_Legendre}
  c_1 &\mapsto e_1 
  = \infty 
  \, , 
  \quad 
  c_2 \mapsto e_2 
  = 0 
  \, , 
  \quad 
  c_3 \mapsto e_3 
  = \lambda 
  \, , 
  \quad 
  c_4 \mapsto e_4 
  = 1 
  \, , \nonumber
  \\
  c_i &\mapsto e_i =\frac{c_{14} c_{2i}}{c_{1i} c_{24}} \,, \; (i=5,\cdots,n)
  \,,
  \quad
  c_{n+1} 
  = \infty \mapsto e_{n+1} \equiv e_{\infty} 
  = \frac{c_{14}}{c_{24}} 
  \, .
\end{align}

We can now state the explicit form for the integrands $\phi_i^{\text{can}}$ of the canonical basis $I^{\text{can}}_i$ for $i = 1, \cdots,n-1$:
\begin{subequations}
  \label{canbasis}
  \begin{align}
    \phi_1^{\rm can}
    & = \frac{1}{\varpi_1 (\lambda)} \frac{\sqrt{c_{13} c_{24}}}{\sqrt{P_4 (z)}} {\rm d} z
    \, , 
    \\
    \phi_{i-3}^{\rm can}
    & = \left(\frac{1}{z - c_i} - \frac{1}{c_{1i}}\right) \frac{\sqrt{P_4 (c_i)}}{\sqrt{P_4 (z)}} {\rm d} z - \frac{\vartheta_1 (e_{i}, \lambda)}{\varpi_1 (\lambda)} \frac{\sqrt{c_{13} c_{24}}}{\sqrt{P_4 (z)}} {\rm d} z
    \, , \quad (i = 5, \cdots, n)
    \, , 
    \\
    \phi_{n-2}^{\rm can}
    & = - \frac{z - c_1}{\sqrt{P_4 (z )}} {\rm d} z - \frac{\vartheta_1 (e_{n+1}, \lambda)}{\varpi_1 (\lambda)} \frac{\sqrt{c_{13} c_{24}}}{\sqrt{P_4 (z)}} {\rm d} z
    \, ,
    \\
    \phi_{n - 1}^{\rm can}
    & = - \frac{1 + 2 \beta_1 \varepsilon}{8 \varepsilon} \varpi_1 (\lambda) \frac{c_{41}(z - c_2)}{c_{42}(z - c_1)} \frac{\sqrt{c_{13} c_{24}}}{\sqrt{P_4 (z )}} {\rm d} z
    - \frac{1}{4} \sum_{i = 5}^{n} \beta_i \vartheta_1 (e_i, \lambda) \left(\frac{1}{z - c_i} - \frac{1}{c_{1i}}\right) \frac{\sqrt{P_4 (c_i)}}{\sqrt{P_4 (z)}} {\rm d} z
    \nonumber
    \\
    & + \frac{1}{4} \beta_{n+1} \vartheta_1 (e_{n+1}, \lambda) \frac{z - c_1}{\sqrt{P_4 (z)}} {\rm d} z - \frac{1}{8 \varepsilon \varpi_1 (\lambda)} \Bigg\{\lambda \varpi_1 (\lambda) [2 (1 - \lambda) \varpi_1^\prime (\lambda) - \varpi_1 (\lambda)] 
    \nonumber
    \\
    & \hspace{10em} + \varepsilon \sum_{i = 2}^{n + 1} \beta_i (1 + \lambda - e_i) \varpi_1 (\lambda)^2 - \varepsilon \sum_{i = 5}^{n + 1} \beta_i \vartheta_1 (e_i, \lambda)^2\Bigg\} \frac{\sqrt{c_{13} c_{24}}}{\sqrt{P_4 (z)}} {\rm d} z
    \, ,
  \end{align}
\end{subequations}
where $\varpi_1$ and $\vartheta_1$ are related to the complete elliptic integrals of the first and the third kinds, respectively:
\begin{subequations}
  \begin{align}
    \varpi_1 (\lambda) 
    & \equiv \frac{2 i}{\pi} \int \limits_{\lambda}^1 \frac{{\rm d} x}{\sqrt{P_L (x)}}
    = \frac{4}{\pi} K (1 - \lambda) 
    \, , 
    \\
    \vartheta_1 (e_i, \lambda)
    & \equiv - \frac{2 i \sqrt{P_L (e_i)} }{\pi e_i} \int \limits_{\lambda}^1 \frac{1}{1- x / e_i} \, \frac{{\rm d} x}{\sqrt{P_L(x)}}
    = - \frac{4 \sqrt{P_L (e_i)}}{\pi e_i} \left[K (1 - \lambda) + \frac{\lambda}{e_i - \lambda} \Pi \left(\frac{e_i (1  - \lambda)}{e_i - \lambda} , 1 - \lambda\right)\right]
    \, .
\end{align}
\end{subequations}

In later derivations of Eq.~\eqref{canbasis}, we will show that the differential equations of $\{I_i^{\text{can}}\}$ are $\varepsilon$-factorized under maximal cut. However, $\varepsilon$-factorization is only a necessary condition for a canonical basis. In the polylogarithmic case, we would impose further requirements besides $\varepsilon$-factorization, to ensure that the canonical integrals evaluate to pure functions. One such requirement is that the integrals should degenerate to UT constants at certain kinematic boundary points. In the elliptic case, one may expect similar properties from a canonical basis \cite{Frellesvig:2023iwr}. In particular, in the kinematic limits where the elliptic curves degenerate to Riemann spheres, one would expect that the canonical integrals degenerate to UT polylogarithmic functions.

With this in mind, it is interesting to investigate the behavior of the basis \eqref{canbasis} in the limit $c_1 \to c_2$, where $\lambda \to 1$ and the elliptic curve degenerates. One can show that the integrands $\phi_i^{\text{can}}$ degenerate to ${\rm d} \log$-forms in the limit $c_1 \to c_2$:
\begin{subequations}
  \label{candeglim}
  \begin{align}
    \lim_{c_1 \to c_2} \phi_1^{\rm can}
    & = - \frac{1}{2} \, {\rm d} \log \left(\frac{1 + \sqrt{\frac{c_{41} (z - c_3)}{c_{31} (z - c_4)}}}{1 - \sqrt{\frac{c_{41} (z - c_3)}{c_{31} (z - c_4)}}}\right) 
    \, ,
    \\
    \lim_{c_1 \to c_2} \phi_{i - 3}^{\rm can}
    & = {\rm d} \log \left(\frac{1 + \sqrt{\frac{c_{i4} (z - c_3)}{c_{i3} (z - c_4)}}}{1 - \sqrt{\frac{c_{i4} (z - c_3)}{c_{i3} (z - c_4)}}}\right) - {\rm d} \log \left(\frac{1 + \sqrt{\frac{c_{41} (z - c_3)}{c_{31} (z - c_4)}}}{1 - \sqrt{\frac{c_{41} (z - c_3)}{c_{31} (z - c_4)}}}\right) 
    \, ,
    \\
    \lim_{c_1 \to c_2} \phi_{n - 2}^{\rm can}
    & = {\rm d} \log \left(\frac{1 + \sqrt{\frac{z - c_3}{z - c_4}}}{1 - \sqrt{\frac{z - c_3}{z - c_4}}}\right) - {\rm d} \log \left(\frac{1 + \sqrt{\frac{c_{41} (z - c_3)}{c_{31} (z - c_4)}}}{1 - \sqrt{\frac{c_{41} (z - c_3)}{c_{31} (z - c_4)}}}\right) 
    \, ,
    \\
    \lim_{c_1 \to c_2} \phi_{n - 1}^{\rm can}
    & = \frac{1}{4} (2 \beta_1 + 2 \beta_2 + \beta_3 + \beta_4) \, {\rm d} \log \left(\frac{1 + \sqrt{\frac{c_{41} (z - c_3)}{c_{31} (z - c_4)}}}{1 - \sqrt{\frac{c_{41} (z - c_3)}{c_{31} (z - c_4)}}}\right) + \frac{1}{2} \sum_{i = 5}^{n + 1} \beta_i \, {\rm d} \log \left(\frac{1 + \sqrt{\frac{c_{i4} (z - c_3)}{c_{i3} (z - c_4)}}}{1 - \sqrt{\frac{c_{i4} (z - c_3)}{c_{i3} (z - c_4)}}}\right)
    \, .
  \end{align}
\end{subequations}
Since ${\rm d} \log$-form integrands lead to UT integrals, we see that the basis in Eq.~\eqref{canbasis} is indeed a natural generalization of the polylogarithmic canonical basis to elliptic cases.

The explicit expressions of the canonical basis \eqref{canbasis}, together with the M\"obius transformation \eqref{M\"obius}, are the main results of this work. They are generic enough to be directly applied to any elliptic integral family with an arbitrary number of branch points. In the following subsections, we show how the above results are derived. In the next section, we apply the above construction to several examples, presenting novel results not available in the literature, to demonstrate the power of our method.

\subsection{The construction of the pre-canonical basis with the Legendre normal form}

\label{subsec_construction}

To derive the canonical basis \eqref{canbasis}, we start from a pre-canonical basis for the integrand after the M\"obius transformation \eqref{M\"obius}. The transformed $u$ function is given by
\begin{equation}
  \label{legendre_notation}
  u_L(x) = [P_L (x)]^{-1/2} \prod_{i = 2}^{n+1} (x - e_i)^{- \beta_i \varepsilon} \equiv [P_L (x)]^{-1/2} \, \bar{u}_L(x)
  \, ,
\end{equation}
where
\begin{equation}
  \beta_{n+1} = - \sum_{i=1}^{n} \beta_i \,.
\end{equation}
The integrals in this sector under maximal cut then have the general form
\begin{equation}
  \label{eq:legendre_family}
  I 
  = \int_{\cal C} u_L(x) \, \varphi  = \int_{\cal C} \bar{u}_L(x) \, \phi  
  \, , 
\end{equation}
where as before, we have absorbed the factor $[P_L (x)]^{-1/2}$ into $\phi$ for later convenience.

To find suitable forms of $\phi$, we take hints from the three kinds of standard elliptic integrals. The incomplete elliptic integral of the first kind is defined as
\begin{equation}
  \label{incomplete_F}
  F (\theta, \lambda)
  = \int \limits_0^{\sin \theta} \frac{{\rm d} t}{\sqrt{(1 - t^2)(1 - \lambda t^2)}} 
  = \frac{1}{2} \int \limits_0^{\lambda \sin^2 \theta} \frac{{\rm d} x}{\sqrt{P_L(x)}} 
  \, ,
\end{equation}
where the differential is holomorphic. For the second equal-sign in the above formula, we have used
\begin{align}
  \label{Legendre2Jacobi}
  \frac{{\rm d} x}{\sqrt{P_L(x)}} = \frac{2 \, {\rm d} t}{\sqrt{(1 - t^2)(1 - \lambda t^2)}}
  \, .
\end{align}
The incomplete elliptic integral of the second kind is defined as
\begin{equation}
  \label{incomplete_E}
  E (\theta, \lambda)
  = \int \limits_0^{\sin \theta} \frac{(1 - \lambda t^2) {\rm d} t}{\sqrt{(1 - t^2)(1 - \lambda t^2)}} 
  = \frac{1}{2} \int \limits_0^{\lambda \sin^2 \theta} \frac{(1 - x) {\rm d} x}{\sqrt{P_L(x)}} 
  \, ,
\end{equation}
where the differential has a double pole at infinity with a vanishing residue. The third kind is defined as 
\begin{equation}
  \Pi (n, \theta, \lambda)
  = \int \limits_0^{\sin \theta} \frac{1}{1 - n \, t^2} \, \frac{{\rm d} t}{\sqrt{(1 - t^2)(1 - \lambda t^2)}}
  = \frac{1}{2} \int \limits_0^{\lambda \sin^2 \theta} \frac{1}{1- n \, x / \lambda} \, \frac{{\rm d} x}{\sqrt{P_L(x)}} 
  \, ,
\end{equation}
with the parameter $n$ characterizing the simple poles of the differential at $t = \pm 1/{\sqrt{n}}$ or $x = \lambda/n$.
The three kinds of complete elliptic integrals are given by the incomplete ones with $\theta = \pi/2$. They're 
\begin{align}
  K (\lambda)
  = F \left(\frac{\pi}{2}, \lambda\right)
  , \quad
  E (\lambda)
  = E \left(\frac{\pi}{2}, \lambda\right)
  , \quad
  \Pi (n, \lambda)
  = \Pi \left(n, \frac{\pi}{2}, \lambda\right)
  \, ,
\end{align}
respectively.

We now propose a set of candidates for the pre-canonical basis, motivated by the three kinds of standard elliptic integrals:\footnote{Note that given the twist in Eq.~\eqref{legendre_notation}, the dimension of the twisted cohomology group is $n-1$, which is the number of solutions to the equation $\mathrm{d} u = 0$. This corresponds to the number of master integrals in this sector under IBP relations without taking into account possible symmetry relations. This counting is independent of the specific underlying geometry.}
\begin{subequations}
  \label{prebasis}
  \begin{align}
    \phi_1^{\text{pre}} & 
    = \frac{{\rm d} x}{\sqrt{P_L(x)}}
    = 2 \, {\rm d} F \left(\sin^{- 1} \sqrt{\frac{x}{\lambda}}, \lambda\right) 
    \, , 
    \\
    \phi_{i - 3}^{\text{pre}} & 
    = \frac{1}{x - e_i} \frac{\sqrt{P_L (e_i)}}{\sqrt{P_L (x)}} {\rm d} x
    = - \frac{2 \sqrt{P_L (e_i)}}{e_i} \, {\rm d} \Pi \left(\frac{\lambda}{e_i}, \sin^{- 1} \sqrt{\frac{x}{\lambda}}, \lambda\right) 
    \, ,       
    \\
    \phi_{n - 1}^{\text{pre}} &
    = \frac{x {\rm d} x}{\sqrt{P_L (x)}}
    = 2 \, {\rm d} F \left(\sin^{- 1} \sqrt{\frac{x}{\lambda}}, \lambda\right) - 2 \, {\rm d} E \left(\sin^{-1} \sqrt{\frac{x}{\lambda}}, \lambda\right) 
    \, ,
  \end{align}
\end{subequations}
where $i = 5, \cdots, n + 1$. The first and the last candidates involve differentials of the first and second kinds of standard elliptic integrals, while the rest involve the third kind. Plugging these $\phi_j^{\text{pre}}$ into Eq.~\eqref{eq:legendre_family}, we get the corresponding master integrals $I_j^{\text{pre}}$. Note we do not introduce any elliptic integrals into the integrands at this step.

We can now derive the differential equations under maximal cut for the pre-canonical master integrals with respect to $\lambda$ and $e_i$ for $i = 5, \cdots, n + 1$. We find that they are in a linear form:
\begin{equation}
  \label{special_linear}
    {\rm d} \vec{I}^{\text{pre}} 
    = \bm{A} \, \vec{I}^{\text{pre}}
    = \left({\bm{A}^{(0)}} + \varepsilon {\bm{A}^{(1)}}\right) \vec{I}^{\text{pre}} 
    \, ,
\end{equation}
where the matrices $\bm{A}^{(0)}$ and ${\bm{A}^{(1)}}$ are independent of $\varepsilon$. To convert the above differential equations to the $\varepsilon$-factorized form, we only need to rotate the matrix $\bm{A}$ by a linear transformation of the master integrals $\vec{I}^{\text{pre}}$, such that the $\varepsilon^0$-part is absent. Before doing that, we may have a look at the explicit form of $\bm{A}^{(0)}$. We write $\bm{A}^{(0)} = \bm{A}^{(0)}_\lambda \mathrm{d}\lambda + \sum_i \bm{A}^{(0)}_{e_i} \mathrm{d}e_i$, where
\begin{equation}
  \label{lambda_de_0}
  \bm{A}^{(0)}_\lambda =
  \begin{pmatrix}
    \frac{1}{2 (1 - \lambda)} & 0 & \cdots & 0 & - \frac{1}{2 \lambda (1 - \lambda)}    
    \\
    - \frac{e_5 (e_5 - 1)}{2 (1 - \lambda) \sqrt{P_L (e_5)}} & 0 & \cdots & 0 & \frac{e_5 (e_5 - 1)}{2 \lambda \left(1 - \lambda\right) \sqrt{P_L (e_5)}}
    \\
    \vdots & \vdots & \ddots & \vdots & \vdots
    \\
    - \frac{e_{n+1} (e_{n+1} - 1)}{2 (1 - \lambda) \sqrt{P_L (e_{n+1})}} & 0 & \cdots & 0 & \frac{e_{n+1} (e_{n+1} - 1)}{2 \lambda (1 - \lambda) \sqrt{P_L (e_{n+1})}} 
    \\
    \frac{1}{2 (1 - \lambda)} & 0 & \cdots & 0 & - \frac{1}{2 (1 - \lambda)}
  \end{pmatrix}
  \, ,
\end{equation}
and
\begin{equation}
  \label{ei_de_0}
  \bm{A}^{(0)}_{e_i} =
  \begin{pmatrix}
    0 & 0 & \cdots & 0 & 0 
    \\
    \vdots & \vdots & \ddots & \vdots & \vdots
    \\
    - \frac{e_i}{2 \sqrt{P_L (e_i)}} & 0 & \cdots & 0 & \frac{1}{2 \sqrt{P_L (e_i)}} 
    \\
    \vdots & \vdots & \ddots & \vdots & \vdots
    \\
    0 & 0 & \cdots & 0 & 0 
  \end{pmatrix}
  \, , \quad (i = 5,\cdots,n+1) \,.
\end{equation}
Notice $\bm{A}^{(0)}$ is special where only the first and the last columns are non-vanishing. This is a direct consequence of the specific choice of the pre-canonical basis in Eq.~\eqref{prebasis}. This fact will be important when we consider the sub-sector dependence of the canonical basis.

\subsection{The transformation to the canonical basis}

\label{subsec_canonical}

We now want to turn the pre-canonical integrands $\vec{\phi}^{\text{pre}}$ in \eqref{prebasis} into a canonical one by acting onto it a transformation matrix ${\cal T}_{\rm can}$. We will perform the transformation in two steps. In other words, we split the matrix ${\cal T}_{\rm can}$ into two parts:
\begin{equation}
  \label{T_can}
  {\cal T}_{\rm can}
  = {\cal T} \, {\cal T}'
  \, .
\end{equation}
The matrix ${\cal T}'$ changes only the first and the last integrals. It is related to the periods of the elliptic curve, defined as the pairings between the differential forms and the integration contours:
\begin{equation}
  \label{periods}
  \varpi_0 (\lambda) 
  \equiv \frac{2}{\pi} \int \limits_0^{\lambda} \frac{{\rm d} x}{\sqrt{P_L (x)}}
  = \frac{4}{\pi} K (\lambda) 
  \, ,
  \quad 
  \varpi_1 (\lambda) 
  \equiv \frac{2 i}{\pi} \int \limits_{\lambda}^1 \frac{{\rm d} x}{\sqrt{P_L (x)}}
  = \frac{4}{\pi} K (1 - \lambda) 
  \, ,
\end{equation}
where $\varpi_0 (\lambda)$ and $\varpi_1 (\lambda)$ are holomorphic around degenerate points $\lambda = 0$ and $\lambda = 1$, respectively. The $1/\pi$ factors are introduced to make the periods of \textit{uniform transcendental weight}-$0$ around degenerate points, while the $i$ factor in the second period is to make the periods real.\footnote{The convention here is slightly different from many literatures.}

The periods are solutions to the Picard-Fuchs equation satisfied by the first pre-canonical integral $I_1^{\text{pre}}$. It is therefore natural to normalize $I_1^{\text{pre}}$ by one of the periods. Here we choose $\varpi_1 (\lambda)$, which is holomorphic as $\lambda \to 1$. As for $I_{n-1}^{\text{pre}}$, we will rotate it to the derivative of the normalized $I_1^{\text{pre}}$ with respect to the modular variable $\tau$ defined as (in the region around $\lambda \to 1$):
\begin{equation}
  \tau 
  \equiv \frac{i \varpi_0 (\lambda)}{\varpi_1 (\lambda)} 
  = \frac{i K (\lambda)}{K (1 - \lambda)} 
  \in {\mathbb H} 
  \, , 
\end{equation}
where $\mathbb{H}$ is the upper-half complex plane. The above strategy is similar to that of \cite{Pogel:2022vat}, which is also equivalent to \cite{Gorges:2023zgv}. Nevertheless, we work with a generic elliptic family in the Legendre normal form rather than a particular Feynman integral family, and our goal is to derive the generic result \eqref{canbasis} of the canonical basis after a M\"obius transformation.

In summary, the non-vanishing elements of the matrix ${\cal T}'$ are given by
\begin{subequations}
  \label{T_der_elements}
  \begin{align}
    ({\cal T}')_{1, 1} 
    & = \frac{1}{\varpi_1 (\lambda)} 
    \, ,
    \\
    ({\cal T}')_{n - 1, n - 1} 
    & = - \frac{1 + 2 \beta_1 \varepsilon}{8 \varepsilon} \varpi_1 (\lambda)
    \, ,
    \\
    ({\cal T}')_{n - 1, j} 
    & = \frac{1}{4} \beta_{j + 3} \frac{\sqrt{P_L (e_{j + 3})}}{e_i - \lambda} \varpi_1 (\lambda)
    \, , \quad   ({\cal T}')_{j,j} = 1
    \,, \quad (2 \leq j \leq n - 2) \,,
    \\
    ({\cal T}')_{n - 1, 1}
    & = \frac{1}{8 \varepsilon} \left[\lambda \varpi_1 (\lambda) - 2 \lambda (1 - \lambda) \varpi_1^\prime (\lambda) + 2 \varepsilon \sum_{i = 2}^{n + 1} \beta_i (e_i - 1) \varpi_1 (\lambda)\right]
    \, .
  \end{align}
\end{subequations}
The matrix transforms $I_{n - 1}^{\text{pre}}$ into
\begin{equation}
  \label{precan_new}
  \left( {\cal T}' \vec{I}^{\text{pre}} \right)_{n - 1}
  = \frac{1}{\varepsilon} \frac{\varpi_1 (\lambda)^2}{\pi W_\lambda} \partial_\lambda \bigg[\frac{1}{\varpi_1 (\lambda)} I_1^{\text{pre}} \bigg] 
  \, , 
\end{equation}
where $W_\lambda$ is the \textit{Wronskian} of $\varpi_1$ and $\varpi_0$ with respect to $\lambda$
\begin{equation}
  \label{wronskian}
  W_\lambda
  = \varpi_1 \partial_\lambda \varpi_0 - \varpi_0 \partial_\lambda \varpi_1 
  = \frac{4}{\pi \lambda (1 - \lambda)}
  \, .
\end{equation}

After imposing ${\cal T}'$, the connection matrix in the differential equations of ${\cal T}' \vec{I}^{\text{pre}}$ is still in a linear form as Eq.~\eqref{special_linear}. We denote the $\varepsilon^0$-part of the connection matrix as $\tilde{\bm A}^{(0)} = \tilde{\bm A}^{(0)}_\lambda \mathrm{d}\lambda + \sum_i \tilde{\bm A}^{(0)}_{e_i} \mathrm{d}e_i$, which is strictly lower triangular. The non-zero elements in $\tilde{\bm A}^{(0)}_{e_i}$ are given by\footnote{The structure of the matrix $\tilde{\bm A}^{(0)}_\lambda$ is similar, albeit with more complicated entries. We do not show its entries here explicitly.}
\begin{subequations}
  \label{ei_de_0_new}
  \begin{align}
    (\tilde{\bm A}^{(0)}_{e_i})_{i - 3, 1}
    & = - \frac{(e_i - \lambda) \varpi_1 (\lambda) + 2 \lambda (1 - \lambda) \varpi_1^\prime (\lambda)}{2 \sqrt{P_L (e_i)}}
    \, ,
    \\
    (\tilde{\bm A}^{(0)}_{e_i})_{n - 1, 1}
    & = - \frac{\beta_i \lambda (1 - \lambda) \varpi_1 (\lambda) \varpi_1^\prime (\lambda)}{4 (e_i - \lambda)}
    \, ,
    \\
    (\tilde{\bm A}^{(0)}_{e_i})_{n - 1, i - 3}
    & = \frac{\beta_i \lambda (1 - \lambda) [\varpi_1 (\lambda) - 2 (e_i - \lambda) \varpi_1^\prime (\lambda)]}{8 (e_i - \lambda) \sqrt{P_L (e_i)}}
    \, .
  \end{align}
\end{subequations}
We now need to rotate these entries away with the matrix $\mathcal{T}$ in Eq.~\eqref{T_can}.

We first construct the matrix element that removes $(\tilde{\bm A}^{(0)}_{e_i})_{i - 3, 1}$. It can be obtained by solving the differential equation
\begin{equation}
  \label{elliptictheta1_partial}
  \frac{\partial \vartheta_1 (e_i, \lambda)}{\partial e_i} 
  = (\tilde{\bm A}^{(0)}_{e_i})_{i - 3,1}
  = - \frac{(e_i - \lambda) \varpi_1 (\lambda) + 2 \lambda (1 - \lambda) \varpi_1^\prime (\lambda)}{2 \sqrt{P_L (e_i)}} 
  \, .
\end{equation}
We denote the function as $\vartheta_1$ to indicate that it is holomorphic around $\lambda = 1$, similar to $\varpi_1 (\lambda)$.
It turns out that the solution can be expressed with the complete elliptic integral of the third kind~\cite{Gorges:2023zgv,Becchetti:2025rrz,Becchetti:2025oyb}:\footnote{By directly integrating $(\tilde{\bm A}^{(0)}_{e_i})_{i - 3,1}$, it is also clear that $\vartheta_1 (e_i, \lambda)$ can be expressed in terms of the complete and incomplete elliptic integrals of the first and second kinds. This indicates a non-trivial relation among elliptic integrals, which might be provable with the \textit{twisted Riemann bilinear relations}~\cite{Duhr:2024rxe}.}
\begin{equation}
  \label{elliptictheta1}
  \vartheta_1 (e_i, \lambda) 
  = - \frac{4 \sqrt{P_L (e_i)}}{\pi e_i} \left[K (1 - \lambda) + \frac{\lambda}{e_i - \lambda} \Pi \left(\frac{e_i (1  - \lambda)}{e_i - \lambda} , 1 - \lambda\right)\right]
  \, ,
\end{equation}
Although we have constructed $\vartheta_1$ from $\tilde{\bm A}^{(0)}_{e_i}$, it also removes the relevant entries in $\tilde{\bm A}^{(0)}_{\lambda}$, without the need to add further $\lambda$-dependent terms.

We still need to remove the entries in the last row of $\tilde{\bm A}^{(0)}_{e_i}$. We adopt the method of Ref.~\cite{Duhr:2024uid}, by looking for a transformation that leads to a constant \textit{intersection matrix} for the canonical basis. Such a transformation can be easily found with an ansatz in terms of $\varpi_1$ and $\vartheta_1$ functions. As a result, we find that all diagonal elements of $\mathcal{T}$ are equal to 1, and the other non-zero elements are given by
\begin{subequations}
  \label{T_elements}
  \begin{align}
    {\cal T}_{j, 1}
    & = - \vartheta_1 (e_{j + 3}, \lambda)
    \, ,
    \\
    {\cal T}_{n - 1, j} 
    & = - \frac{1}{4} \beta_{j + 3} \left[\vartheta_1 (e_{j + 3}, \lambda) + \frac{e_{j + 3} (e_{j + 3} - 1) \varpi_1 (\lambda)}{\sqrt{P_L (e_{j + 3})}}\right]
    \, ,
    \\
    {\cal T}_{n - 1, 1}
    & = \frac{1}{8} \left[\sum_{i = 5}^{n + 1} \beta_i \vartheta_1 (e_i, \lambda)^2 - \sum_{i = 2}^{n + 1} \beta_i (e_i + \lambda - 1) \varpi_1 (\lambda)^2\right]
    \, ,
  \end{align}
\end{subequations}
where $2\leq j\leq n-2$. We can now combine ${\cal T}'$ from~\eqref{T_der_elements} and ${\cal T}$ from~\eqref{T_elements} into ${\cal T}_{\rm can}$ as in Eq.~\eqref{T_can}. Acting the matrix ${\cal T}_{\rm can}$ onto \eqref{prebasis}, we readily obtain a canonical basis for the integral family in the Legendre normal form.

It is interesting to have a look at the differential equations satisfied by our canonical basis. The connection matrix of the canonical basis has the $\varepsilon$-factorized form
\begin{equation}
  \bm{A}^{\rm can} = {\bm A}_{\lambda}^{\rm can} {\rm d} \lambda + \sum_i {\bm A}_{e_i}^{\rm can} {\rm d} e_i = \varepsilon \bm{A}^{\text{can},(1)} \,.
\end{equation}
We find that the entries in $\bm{A}^{\rm can}$ only involve the transcendental functions $\varpi_{1} (\lambda)$ and $\vartheta_{1} (e_i, \lambda)$, while $\varpi^\prime_{1} (\lambda)$ does not appear. The power of $\varpi_{1} (\lambda)$ in the denominator is at most $2$, and $\vartheta_{1} (e_i, \lambda)$ only appears in the numerator. Furthermore, the matrix at the degenerate point $\lambda = 1$ is UT and has only a simple pole, e.g.,
\begin{equation}
  \lim_{\lambda \to 1} {\bm A}_{\lambda}^{\rm can} {\rm d} \lambda
  = - \varepsilon 
  \begin{pmatrix}
    \frac{\beta_{34}}{2} & 0 & \cdots & 0 & 1 
    \\
    0 & 0 & \cdots & 0 & 0
    \\
    \vdots & \vdots & \ddots & \vdots & \vdots
    \\
    0 & 0 & \cdots & 0 & 0
    \\
    \frac{\beta_{34}^2}{4} & 0 & \cdots & 0 & \frac{\beta_{34}}{2}
  \end{pmatrix} {\rm d} \log (1 - \lambda)
  \, ,
\end{equation}
and similar behaviors hold for ${\bm A}_{e_i}^{\rm can} {\rm d} e_i$ as well. These properties satisfy the conditions for a canonical basis as advocated in \cite{Dlapa:2022wdu}. Moreover, our canonical basis has a constant intersection matrix and thus is self-dual~\cite{Pogel:2024sdi,Duhr:2024xsy} (up to a constant rotation) by construction. Such properties are required in the definition of a canonical basis in~\cite{Duhr:2024uid}. The above discussions, together with those around Eq.~\eqref{candeglim}, indicate that our canonical basis is indeed ``canonical'' in a broad sense.

At this point, we would like to discuss a bit about the sub-sector dependence of our canonical basis. Although our construction works under the maximal cut with no information about the sub-sectors, the simple form of our pre-canonical basis \eqref{prebasis} allows us to naturally promote them to Feynman integrals $\vec{I}^{\text{pre}}$ without any cuts. It turns out that, in the differential equations of many examples, the sub-sector dependence of $\vec{I}^{\text{pre}}$ is rather simple, and can be easily converted to an $\varepsilon$-factorized form. However, this simplicity could be spoiled by the rotation matrix ${\cal T}_{\rm can}$, especially due to its non-trivial $\varepsilon$-dependence. Fortunately, the $\varepsilon$-dependence in ${\cal T}_{\rm can}$ only appears in the last row. It is possible to disentangle the $\varepsilon$-dependence by decomposing ${\cal T}_{\rm can}$ as
\begin{equation}
  {\cal T}_{\rm can}
  = {\cal T}_{\rm can}^{(0)} \, {\cal T}_{\rm can}^{(\varepsilon)}
  \, ,
\end{equation}
where ${\cal T}_{\rm can}^{(0)}$ is $\varepsilon$-independent and
\begin{equation}
  \label{su_eps}
  {\cal T}_{\rm can}^{(\varepsilon)}
  = 
  \begin{pmatrix}
    \frac{1}{\varpi_1 (\lambda)} & 0 & \cdots & 0 & 0
    \\
    0 & 1 & \cdots & 0 & 0
    \\
    \vdots & \vdots & \ddots & \vdots & \vdots
    \\
    0 & 0 & \cdots & 1 & 0
    \\
    \frac{\lambda [\varpi_1 (\lambda) - 2 (1 - \lambda) \varpi_1^\prime (\lambda)]}{8 \varepsilon} & 0 & \cdots & 0 & - \frac{(1 + 2 \beta_1 \varepsilon) \varpi_1 (\lambda)}{8 \varepsilon}
  \end{pmatrix}
  \, .
\end{equation}
We can see that ${\cal T}_{\rm can}^{(\varepsilon)}$ only involves $\varpi_1$ and its derivative, which is relatively easy to handle. Moreover, the $\varepsilon^0$-part in the differential equations of ${\cal T}_{\rm can}^{(\varepsilon)} \vec{I}^{\text{pre}}$ is strictly lower triangular within the top-sector, similar to the case in Eq.~\eqref{ei_de_0_new}. These properties are useful to simplify the sub-sector dependence, leading to a canonical basis for the full integral family. In fact, the simple $\varepsilon$-dependence of ${\cal T}_{\rm can}^{(\varepsilon)}$ often makes it easier to simplify part of sub-sector dependence directly from the pre-canonical basis. We will elaborate on this procedure in the next Section.

\subsection{The M\"obius transformation}

\label{subsec_legendre2original}

As a final ingredient in our construction, we now briefly discuss the M\"obius transformation \eqref{M\"obius}, that takes any univariate elliptic integral defined by the degree-4 polynomial $P_4(z)$ in Eq.~\eqref{eq:general_u} into the Legendre normal form. This leads to our final result \eqref{canbasis} for the canonical basis in a generic elliptic integral family.

It is well-known that M\"obius transformations are automorphisms of the Riemann sphere, or equivalently, $\mathbb{CP}^1$. Two elliptic curves are isomorphic if they are related by a M\"obius transformation.
A M\"obius transformation can be expressed as follows:
\begin{equation}
  z
  \mapsto x 
  = T (z) 
  = \frac{a\, z + b}{c\, z + d}
  \, , 
  \quad
  (a, b, c, d\in \mathbb{C}, \, a d - b c \neq 0)
  \, .
\end{equation}
The inverse transformation is given by
\begin{equation}
  z = T^{-1}(x) = \frac{d\,x-b}{a-c\,x} \,.
\end{equation}
Under a M\"obius transformation $T$, the difference between two points transforms as
\begin{equation}
  \label{difference}
  z_1 - z_2 
  \mapsto x_1 - x_2
  = T (z_1) - T (z_2) 
  = \frac{(a d - b c) (z_1 - z_2)}{(c z_1 + d) (c z_2 + d)}
  \, ,
\end{equation}
while the differential transforms as
\begin{equation}
  \label{differential}
  {\rm d} z 
  \mapsto {\rm d} x 
  = {\rm d} T (z)
  = \frac{(a d - b c) {\rm d} z}{(c z + d)^2} \, .
\end{equation}

For our purpose, we would like to map three out of the four branch points in $P_4(z)$ to $\{0, 1, \infty\}$ and the other one to the modulus $\lambda$. For concreteness, we assume $c_1 < c_2 < c_3 < c_4$, and require
\begin{equation}
  z 
  \mapsto x 
  = T (z) 
  \, , 
  \quad 
  c_1 \mapsto e_1 
  = \infty 
  \, , 
  \quad 
  c_2 \mapsto e_2 
  = 0 
  \, , 
  \quad 
  c_3 \mapsto e_3 
  = \lambda 
  \, , 
  \quad 
  c_4 \mapsto e_4 
  = 1 
  \, .
\end{equation}
The above requirement uniquely fixes the form of the transformation as
\begin{equation}
  \label{mobius_trans}
  x = T (z) 
  = \frac{(z - c_2) c_{14}}{(z - c_1) c_{24}}
  \, .
\end{equation}
Under the transformation, $c_{\infty}$ is mapped to 
\begin{equation}
  c_{\infty} 
  = \infty \mapsto e_{\infty} 
  = \frac{c_{14}}{c_{24}} 
  \, .
\end{equation}
We then have the relations
\begin{align}
  {\rm d} x = \frac{c_{21} c_{41}}{c_{42}} \frac{{\rm d} z}{(z - c_1)^2} 
  \, , \quad 
  x - e_\infty = \frac{c_{21} c_{14}}{c_{42}} \frac{1}{z - c_1}
  \,, \quad
  x - e_i = \frac{c_{21} c_{14}}{c_{42} c_{1i}} \frac{z - c_i}{z - c_1}  \,, 
\end{align}
with $2\leq i\leq n$.
Plugging the above relations into $u_L(x)$, we find
\begin{equation}
  u_L(x) \, \mathrm{d}x \propto u(z) \, \mathrm{d}z \,,
\end{equation}
where we have again suppressed factors that only depend on kinematic variables.

With the relations above, we can already rewrite the canonical basis $\mathcal{T}_{\text{can}} \phi^{\text{pre}}$ obtained in the previous subsection in terms of the variable $z$, leading to Eq.~\eqref{canbasis}. On the other hand, it is interesting to investigate the behavior of the pre-canonical basis in Eq.~\eqref{prebasis} under this transformation. They are given by
\begin{subequations}
  \label{pre_basis_general}
  \begin{align}
    \phi_1^{\text{pre}} 
    & = \sqrt{c_{13} c_{24}} \frac{{\rm d} z}{\sqrt{P_4 (z)}} 
    \, , 
    \\
    \phi_{i - 3}^{\text{pre}} 
    & = \sqrt{P_4 (c_i)} \left(\frac{1}{z - c_i} - \frac{1}{c_{1i}}\right) \frac{{\rm d} z}{\sqrt{P_4 (z)}} \, , \quad (i=5,\cdots,n) \,,
    \\
    \phi_{n - 2}^{\text{pre}} 
    & = - (z - c_1) \frac{{\rm d} z}{\sqrt{P_4 (z )}} 
    \, ,
    \\
    \phi_{n - 1}^{\text{pre}} 
    & = \sqrt{c_{13} c_{24}} \frac{c_{41} (z - c_2)}{c_{42} (z - c_1)} \frac{{\rm d} z}{\sqrt{P_4 (z )}} 
    \, .
  \end{align}
\end{subequations}

In Eq.~\eqref{candeglim}, we have studied the asymptotic behavior of our canonical basis in the degenerate limit $c_1 \to c_2$. Interestingly, our pre-canonical basis also exhibits the nice property of $\mathrm{d}\log$-form behavior in this limit. The first two basis integrals asymptotically approach the same ${\rm d} \log$-form 
\begin{equation}
  \label{phi12deglim}
  \lim_{c_1 \to c_2} \phi_1^{\text{pre}} 
  = \lim_{c_1 \to c_2} \phi_{n - 1}^{\text{pre}} 
  = \frac{\sqrt{c_{13} c_{14}}}{(z - c_1) \sqrt{(z - c_3) (z - c_4)}} {\rm d} z 
  = - {\rm d} \log \left(\frac{1 + \sqrt{\frac{c_{41} (z - c_3)}{c_{31} (z - c_4)}}}{1 - \sqrt{\frac{c_{41} (z - c_3)}{c_{31} (z - c_4)}}}\right) 
  \, .
\end{equation}
Similarly, $\phi_{i - 3}^{\text{pre}}$ and $\phi_{n - 1}^{\text{pre}}$ approach two ${\rm d} \log$-forms as well:
\begin{align}
  \label{phi34deglim}
  \lim_{c_1 \to c_2} \phi_{i - 3}^{\text{pre}} & 
  = \frac{\sqrt{c_{i3} c_{i4}}}{(z - c_i) \sqrt{(z - c_3) (z - c_4)}} {\rm d} z
  = {\rm d} \log \left(\frac{1 + \sqrt{\frac{c_{i4} (z - c_3)}{c_{i3} (z - c_4)}}}{1 - \sqrt{\frac{c_{i4} (z - c_3)}{c_{i3} (z - c_4)}}}\right) 
  \, ,
  \\
  \lim_{c_1 \to c_2} \phi_{n - 2}^{\text{pre}} & 
  = - \frac{{\rm d} z}{\sqrt{(z - c_3) (z - c_4)}}
  = {\rm d} \log \left(\frac{1 + \sqrt{\frac{z - c_3}{z - c_4}}}{1 - \sqrt{\frac{z - c_3}{z - c_4}}}\right) 
  \, .
\end{align}

As a final remark, we emphasize that the construction in this Section has focused on the kinematic region around the degenerate point $c_1 \to c_2$ and thus $\lambda \to 1$. If one is interested in the limit $c_2 \to c_3$ and thus $\lambda \to 0$, a canonical basis can be easily constructed by replacing $\varpi_1$ by $\varpi_0$ in Eq.~\eqref{canbasis}, and replacing $\vartheta_1$ by
\begin{equation}
  \label{elliptictheta0}
  \vartheta_0 (e_i, \lambda) 
  \equiv - \frac{4 \sqrt{P_L (e_i)} \Pi (\frac{\lambda}{e_i}, \lambda)}{\pi e_i} 
  \, ,
\end{equation}
which satisfies the differential equation
\begin{equation}
  \label{elliptictheta0_partial}
  \frac{\partial \vartheta_0 (e_i, \lambda)}{\partial e_i} 
  = - \frac{(e_i - \lambda) \varpi_0 (\lambda) + 2 \lambda (1 - \lambda) \varpi_0^\prime (\lambda)}{2 \sqrt{P_L (e_i)}} 
  \, .
\end{equation}
In the meantime, the M\"obius transformation should be modified accordingly.
The functions $\varpi_0$ and $\vartheta_0$ are holomorphic around $\lambda \to 0$. As a result, one can show that the canonical basis degenerates to $\mathrm{d}\log$-forms in that limit.

\section{Examples}

\label{sec_eg}

\begin{table}
  \centering
  \begin{tabular}{ccccc}
    \hline
    \textbf{Family} & \textbf{Graph} & \textbf{Scales} & \textbf{Sectors} & \textbf{References} 
    \\
    \hline 
    \parbox[c]{14em}{Unequal-mass sunrise (sr)} & \parbox[c]{8em}{\includegraphics[width=3cm]{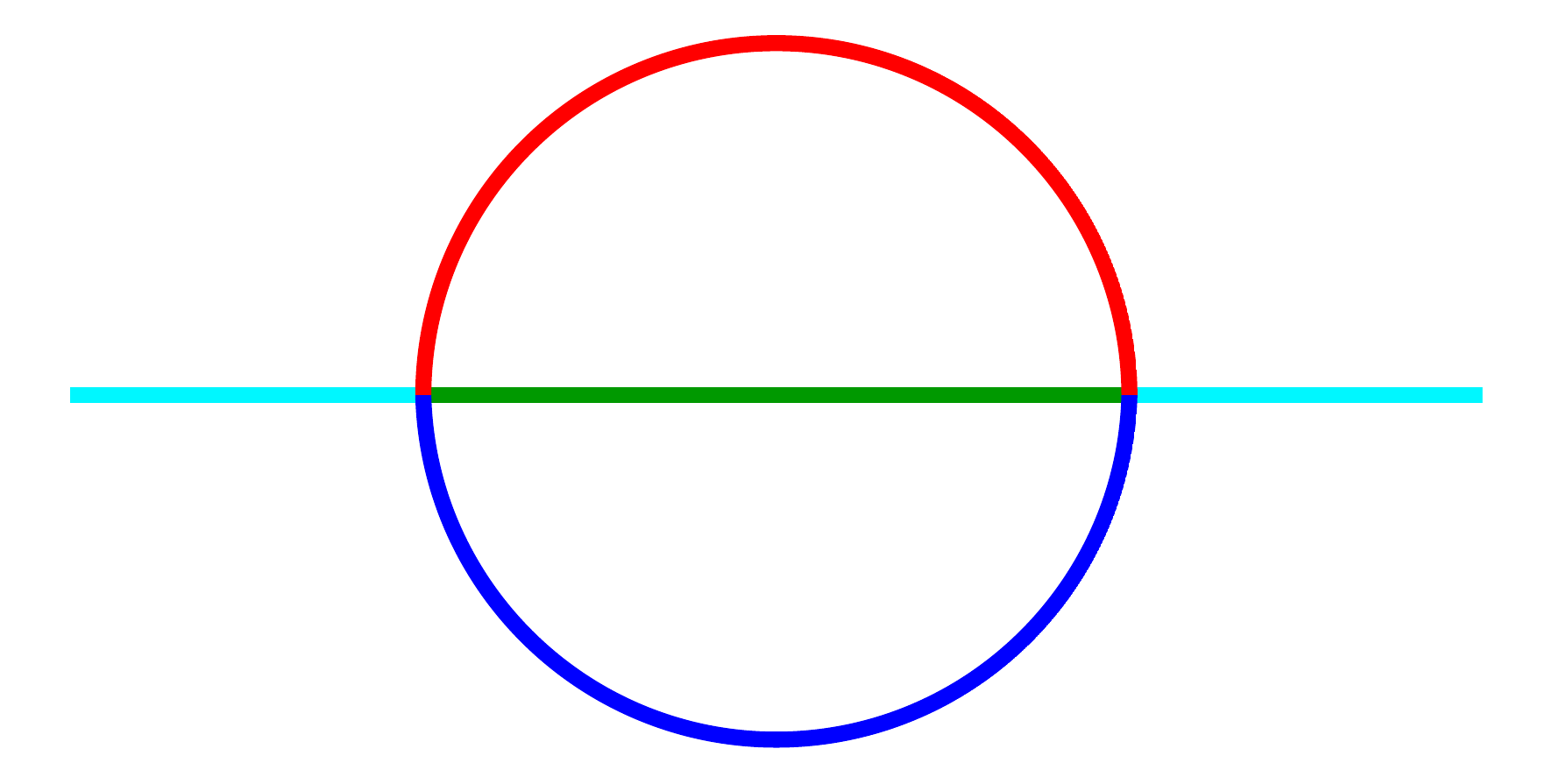}} & $4$ & All~($7$) & \cite{Bogner:2019lfa} 
    \\
    \parbox[c]{14em}{An elliptic sub-sector of $2$-loop $t \bar{t} W$ production (ttW)}  & \parbox[c]{8em}{\includegraphics[width=3cm]{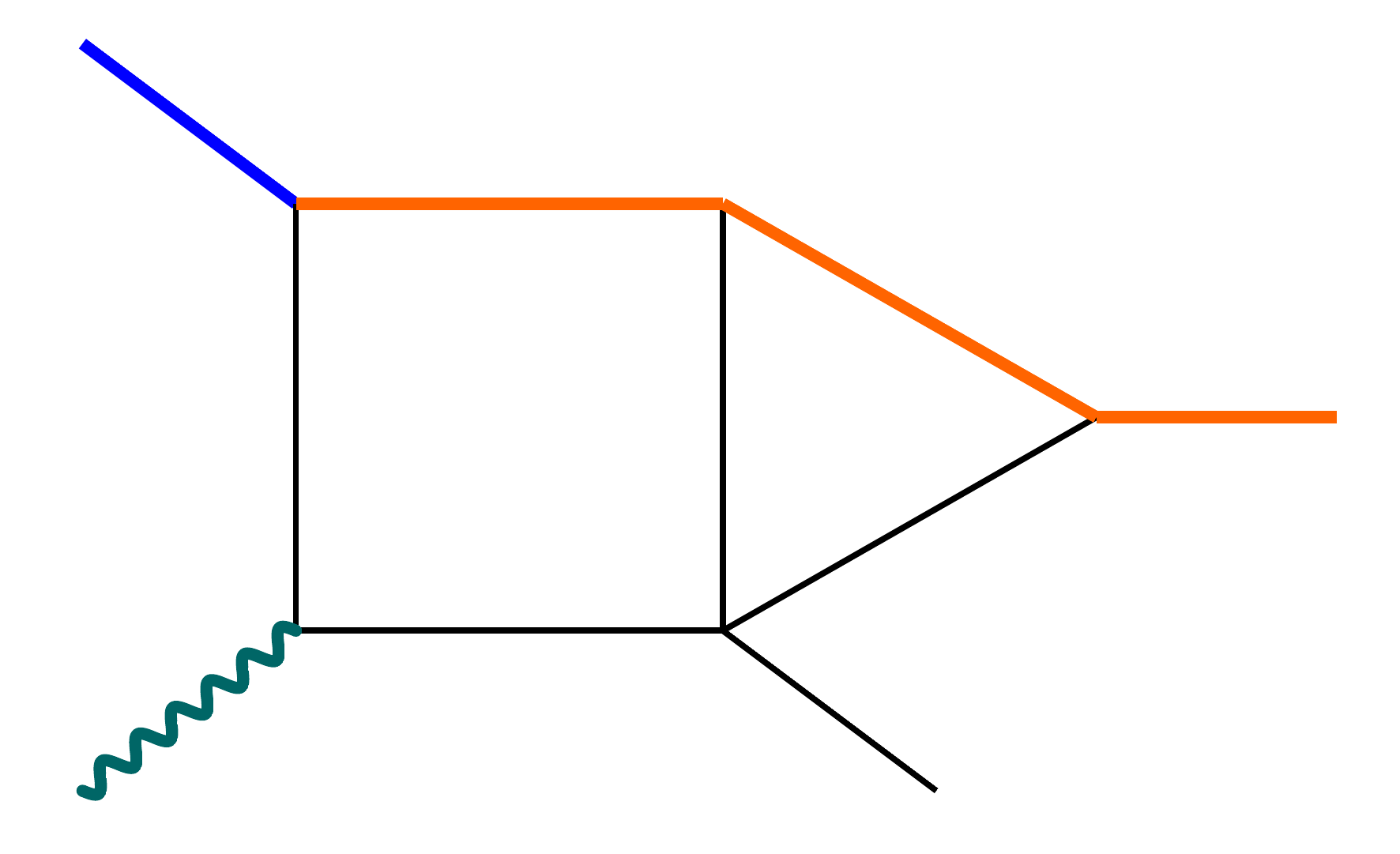}} & $5$ & Top ($3$) & \parbox[c]{3.5em}{New, see also~\cite{Becchetti:2025qlu}}
    \\
    \parbox[c]{14em}{${\cal T}_{3F}$ branch of $2$-loop $W$-pair production via light quark-antiquark annihilation (t3f)} & \parbox[c]{8em}{\includegraphics[width=3cm]{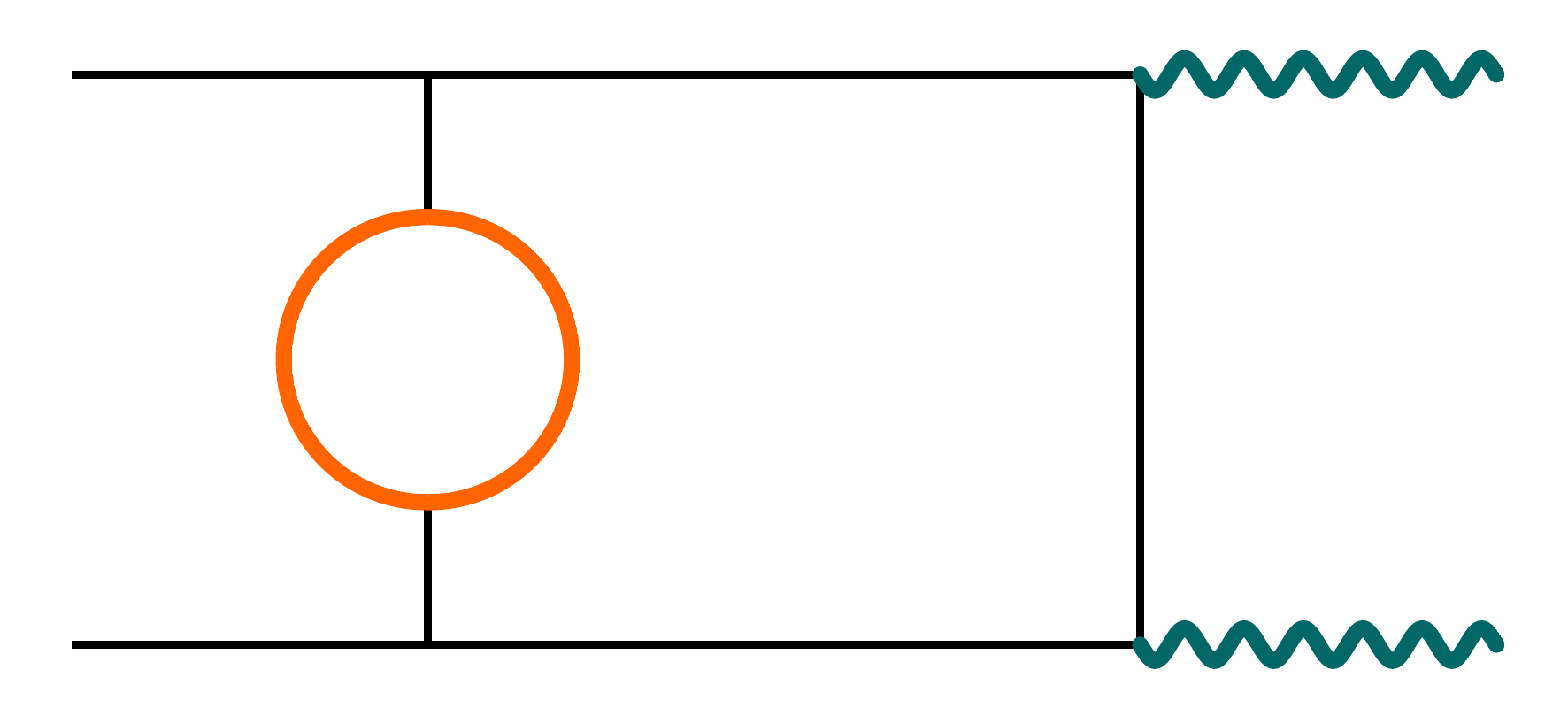}} & $4$ & All~($15$) & \parbox[c]{3.5em}{New, see also~\cite{He:2024iqg}}
    \\
    \parbox[c]{14em}{Non-planar double box A (npdbA)} & \parbox[c]{8em}{\includegraphics[width=3cm]{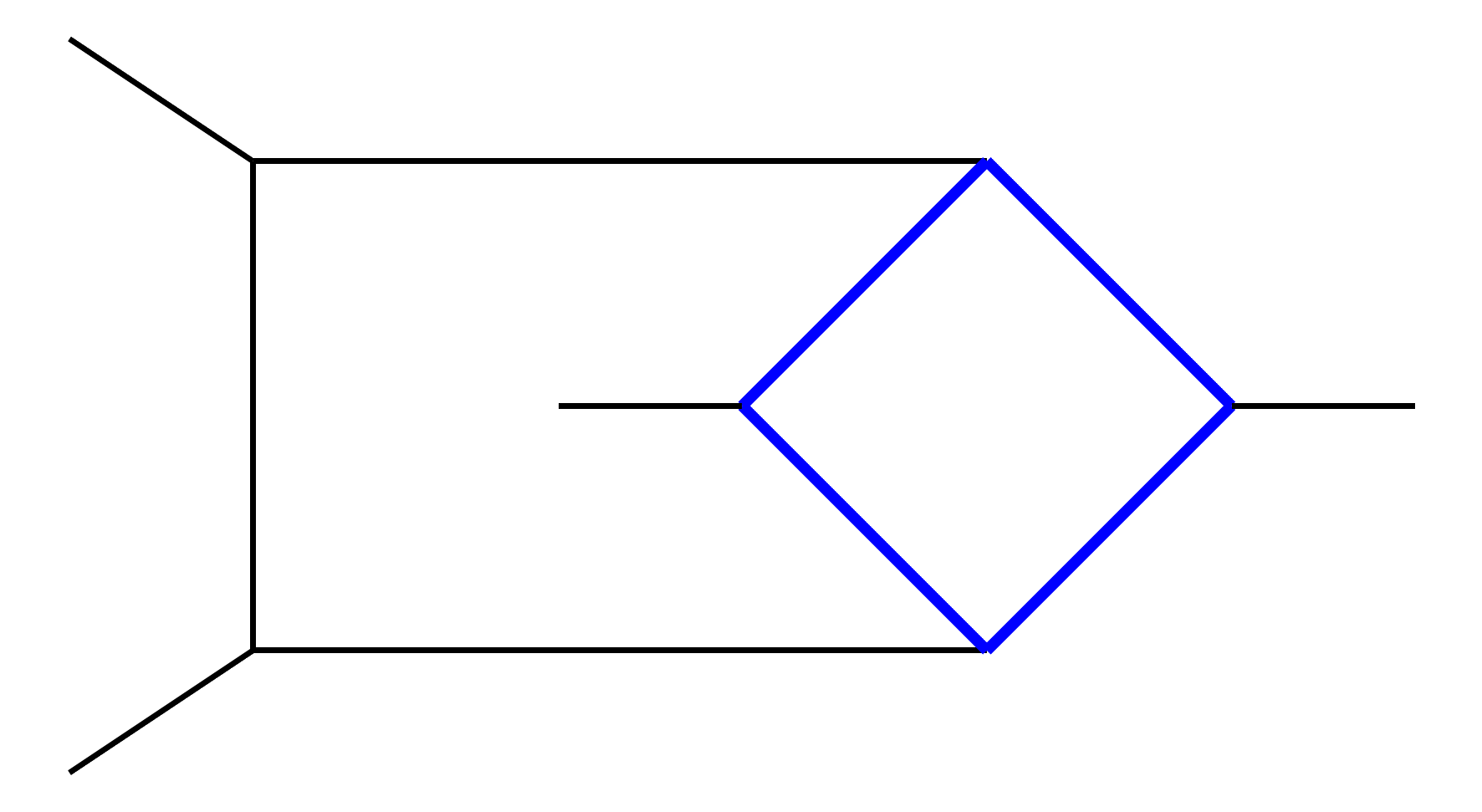}} & $3$ & All~($36$) & \cite{Becchetti:2025rrz} 
    \\
    \parbox[c]{14em}{Non-planar double box B (npdbB)} & \parbox[c]{8em}{\includegraphics[width=3cm]{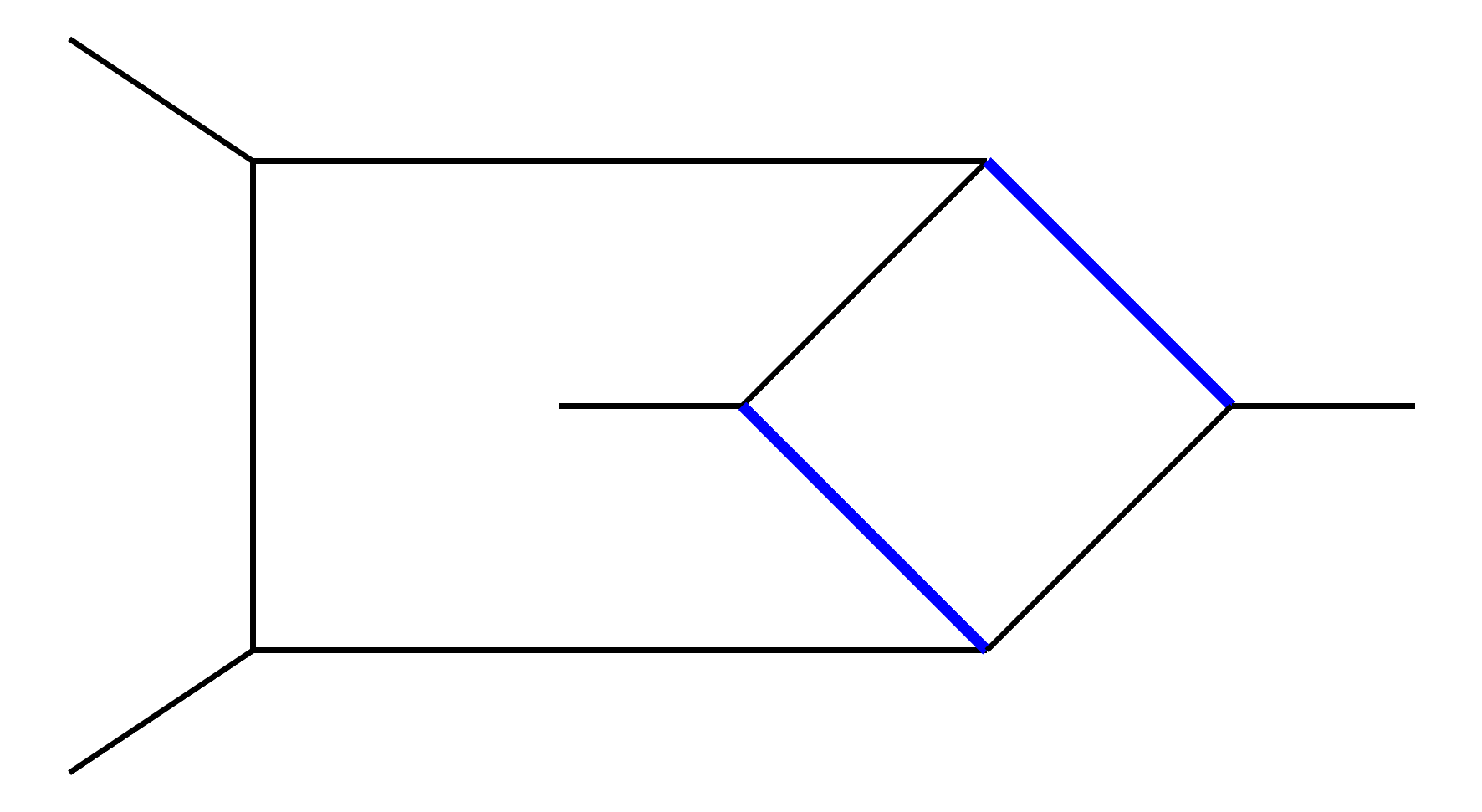}} & $3$ & Top ($4$) $+$ Triangle ($3$) & \cite{Schwanemann:2024kbg} 
    \\
    \hline
  \end{tabular}
  \caption{The examples we will consider are listed in the above table. The first column shows the integral family, with the abbreviated name used in this paper given in brackets. The second column contains the Feynman diagram for the family. Thick, colored lines represent propagators with non-zero invariant masses, and distinct colors represent different masses. The third column gives the number of kinematic scales; the number of independent dimensionless parameters is one less than this value. The fourth column indicates the sectors we consider, where the basis integrals are transformed into canonical form. The number in brackets is the number of master integrals. The last column provides references for the canonical bases of the families. Two of these results are new to this work, and their relevant citations are also given.}
  \label{tab_examples}
\end{table}

In this section, we apply our results to various examples to demonstrate the generality of the canonical integrands. The sunrise families are prototypes of elliptic Feynman integrals, they are well-studied in~\cite{Gorges:2023zgv,Bogner:2019lfa}. We will use the unequal-mass sunrise family as a pedagogical example. We then apply our methods to two phenomenologically interesting integral families, whose canonical bases were not reported in the literature. The first one is relevant to $t \bar{t} W$ production~\cite{Becchetti:2025qlu}, while the second one is relevant to $W$-pair production~\cite{He:2024iqg}. We present their canonical bases for the first time. In particular, for the $W$-pair production case, we derive the canonical basis for the full family without any cuts. This benefits from the simplicity of the sub-sector dependence of our pre-canonical basis.

Two further integral families are discussed as a demonstration of two-variable constructions under next-to-maximal cut. These are non-planar double box families with two different mass configurations. The first configuration appears in di-jet and di-photon production~\cite{Becchetti:2023wev}, as well as Higgs boson pair production in the small-mass limit~\cite{Xu:2018eos,Wang:2020nnr}. It has also been discussed in earlier works~\cite{Ahmed:2024tsg,Becchetti:2025rrz}. The second configuration is relevant to a subset of the NNLO electroweak corrections to M\o{}ller scattering ($e^- e^- \to e^- e^-$). This subset involves the exchange of a single type of massive gauge bosons ($W^\pm$ or $Z^0$), and was recently discussed in \cite{Schwanemann:2024kbg}. For each family, we select a next-to-maximal cuts where two elliptic sectors survive. We show that the sub-sector dependence can be greatly simplified with such two-variable constructions.

For reference, the examples discussed in this section are summarized in Tab.~\ref{tab_examples}.

\subsection{Unequal-mass Sunrise}

\label{subsec_sunrise}

\begin{figure}
  \centering
  \includegraphics[width=7cm]{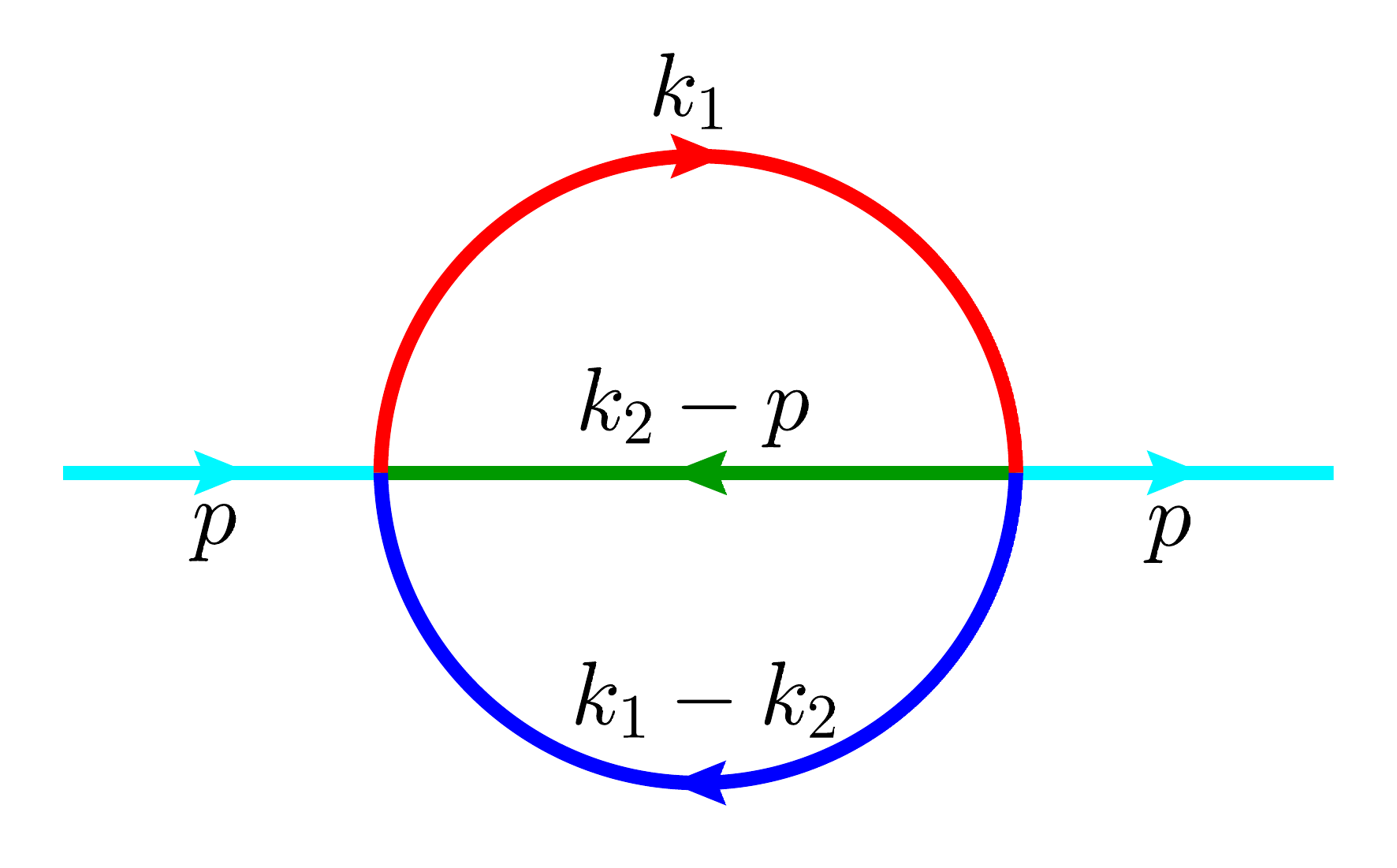}
  \caption{The unequal-mass sunrise diagram is shown above, where different colors represent different masses for the propagators. The only elliptic sector in the family is the top sector. The sub-sectors are obtained by pinching the propagators, which are double tadpoles. The case is well-studied in the literature~\cite{Bogner:2019lfa}.}
  \label{sunrise_fig}
\end{figure}

The unequal-mass sunrise diagram is shown in Fig.~\ref{sunrise_fig}. The external legs are massive, and $D_1$, $D_2$ and $D_3$ are three massive propagators with distinctive masses shown in Fig.~\ref{sunrise_fig}, while $D_4$ and $D_5$ are two ISPs, i.e., $v_4 \leq 0,v_5 \leq 0$. They are defined as
\begin{equation}
  \label{props_sunrise}
  \begin{aligned}
  D_1 
  & = k_1^2 - m_1^2 
  \, , 
  \quad &
  D_2 
  & = (k_2 - p)^2 - m_2^2 
  \, , 
  \quad &
  D_3 
  & = (k_1 - k_2)^2 - m_3^2 
  \, ,
  \\
  D_4 
  & = k_2^2 
  \, , 
  \quad &
  D_5 
  & = (k_1 - p)^2 
  \, .
  \end{aligned}
\end{equation}
The integral family depends on masses $m_i$'s and the only Mandelstam variable associated with the incoming momentum $p$:
\begin{equation}
  s \equiv p^2 \, .
\end{equation}
We have defined the integrals as dimensionless by introducing the scale $\mu^2$. We will choose $\mu^2 = m_1^2$, such that the integrals are functions of the dimensionless variables
\begin{equation}
  \label{sunrise_variables}
  (x + y)^2 
  \equiv \frac{4 s}{m_1^2}
  \, ,
  \quad 
  (x - y)^2
  \equiv \frac{4 m_2^2}{m_1^2}
  \, ,
  \quad
  z^2 
  \equiv \frac{m_3^2}{m_1^2}
  \, .
\end{equation}
The definitions for dimensionless variables are chosen to make the branch points of the corresponding elliptic curves free of square roots, which we will see explicitly soon.

We use \texttt{Kira}~\cite{Klappert:2020nbg} to reduce the integrals in this family to a set of 7 master integrals, chosen as
\begin{equation}
  \label{sunrise_basis}
  \big\{
    I_{11200}
    \, ,
    I_{12100}
    \, ,
    I_{21100}
    \, ,
    I_{11100}
    \, ,
    I_{01100}
    \, ,
    I_{10100}
    \, ,
    I_{11000}
  \big\}
  \, .
\end{equation}
The first $4$ integrals are in the elliptic top sector, while the remaining $3$ sub-sector integrals are non-elliptic. We will study the family in $d = 2 - 2 \varepsilon$, where the non-elliptic master integrals are automatically canonical. We can then focus on the top sector.

The maximal cut of the master integral $I_{11100}$ is given by
\begin{equation}
  \label{maxcut_sunrise}
  {\rm MaxCut}[I_{11100}]
  = \int_{\cal C} \, u (z_4) {\rm d} z_4
  \, ,
\end{equation}
where
\begin{equation}
  \label{u_sunrise}
  u (z_4)
  = [P_4 (z_4)]^{- 1/2} \prod_{i = 1}^5 (z_4 - c_i)^{- \beta_i \varepsilon}
  \, .
\end{equation}
The branch points can be expressed with the dimensionless variables as
\begin{equation}
  \label{branch_sunrise}
  c_1 
  = (z - 1)^2
  \, ,
  \quad
  c_2 
  = x^2
  \, ,
  \quad
  c_3 
  = y^2
  \, ,
  \quad
  c_4 
  = (z + 1)^2
  \, ,
  \quad
  c_5 
  = 0
  \, .
\end{equation}
We consider the region $0 < x < y < 2$, $1 < z < 1 + x$, where the ordering of the branch points is given by $c_5 < c_1 < c_2 < c_3 < c_4$. According to Eqs.~\eqref{lambda} and \eqref{z2_Legendre}, the branch points are mapped into the Legendre normal form as: 
\begin{equation}
  \label{variable_sunrise}
  e_3 = \lambda
  = \frac{4 z (x^2 - y^2)}{[x^2 - (z + 1)^2][y^2 - (z - 1)^2]}
  \, ,
  \quad
  e_5 
  = \frac{4 x^2 z}{(z - 1)^2 [(z + 1)^2 - x^2]}
  \, ,
  \quad
  e_6 = e_\infty
  = \frac{4 z}{(z + 1)^2 - x^2}
  \, .
\end{equation}
Note that in the limit $x \to z - 1$, we have $c_1 \to c_2$ and $\lambda \to 1$.

To write down the canonical basis for the top sector, we just need to plug Eqs.~\eqref{branch_sunrise} and~\eqref{variable_sunrise} into Eq.~\eqref{canbasis}, followed by converting the integrands into linear combinations of the master integrals $I_{11200}$, $I_{12100}$ and $I_{11100}$. Accidentally, the sub-sector dependence of these top-sector canonical basis turns out to be $\varepsilon$-factorized automatically. As we will see in later examples, the sub-sector dependence can be non-trivial in more complicated cases, which requires further manipulation.

\subsection{An elliptic sub-sector of $2$-loop $t \bar{t} W$ production}

\label{subsec_ttW}

\begin{figure}
  \centering
  \includegraphics[width=15cm]{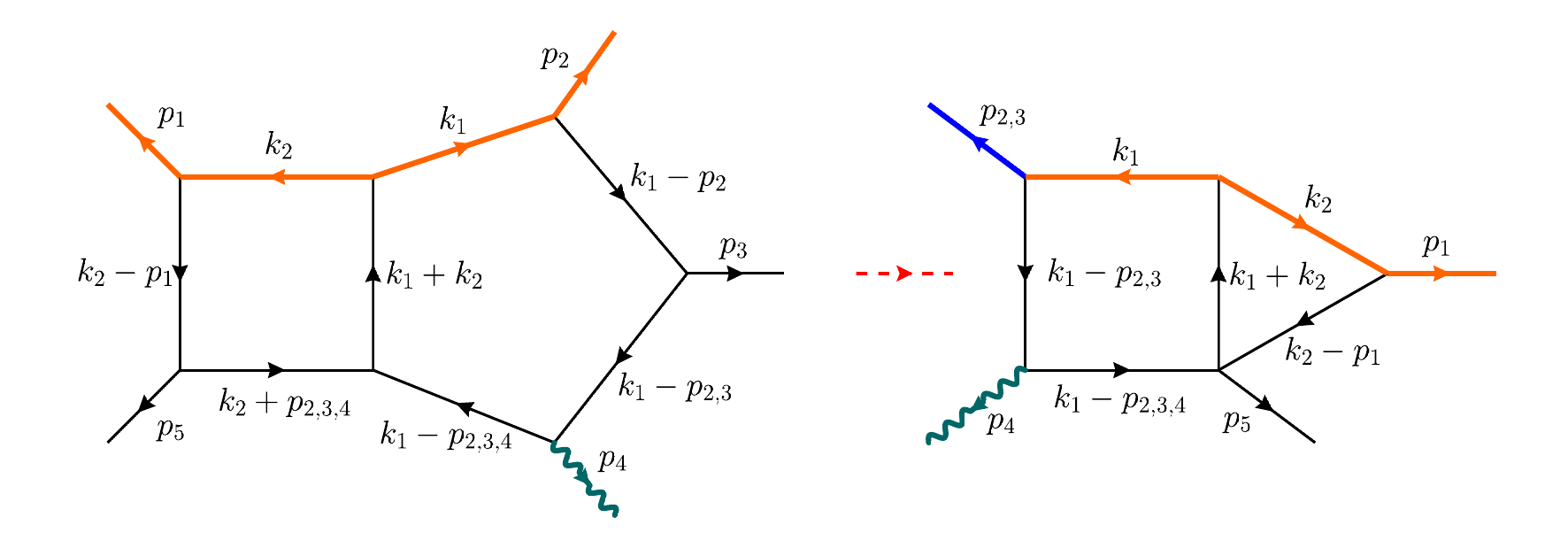}
  \caption{A family contributing to $2$-loop $t \bar{t} W$ production~\cite{Becchetti:2025qlu} is shown in the left panel. There are two four-point elliptic sectors, one of which has 4 scales and its canonical basis has been given in~\cite{Becchetti:2025oyb}. The remaining elliptic sector is shown in the right panel, which is more complicated with 5 scales. In the diagrams, thick orange lines represent top quarks with mass $m_t$, the curvy lines represent $W$ bosons with mass $m_W$, and the thick blue line is an off-shell particle coming from pinching the propagator with momentum flow $k_1 - p_2$ while all the others are massless.}
  \label{ttW_fig}
\end{figure}
The diagram for an elliptic sub-sector of $2$-loop $t \bar{t} W$ production is shown in Fig.~\ref{ttW_fig}. We follow the notations in~\cite{Becchetti:2025qlu} with slight modifications. The propagators are 
\begin{equation}
  \label{ttW_prop}
  \begin{aligned}
    D_1 
    & = k_1^2 - m_t^2
    \, , 
    \quad &
    D_2 
    & = (k_1 - p_2 - p_3)^2
    \, , 
    \quad &
    D_3 
    & = (k_1 - p_2 - p_3 - p_4)^2 
    \, ,
    \\
    D_4 
    & = k_2^2 - m_t^2 
    \, , 
    \quad &
    D_5 
    & = (k_2 - p_1)^2 
    \, ,
    \quad &
    D_6 
    & = (k_1 + k_2)^2 
    \, , 
    \\
    D_7
    & = (k_2 + p_2 + p_3 + p_4)^2 
    \, ,
    \quad &
    D_8
    & = (k_1 + p_1)^2
    \, ,
    \quad &
    D_9 
    & = (k_2 + p_2 + p_3)^2 - m_t^2
    \, .
  \end{aligned}
\end{equation}
The kinematic variables are defined as
\begin{equation}
  \label{ttW_kin}
  s_{45}
  \equiv (p_1 + p_2 + p_3)^2
  \, ,
  \quad 
  s_{14}
  \equiv (p_1 + p_4)^2
  \, ,
  \quad 
  s_{23}
  \equiv (p_2 + p_3)^2
  \, .
\end{equation}
The scale $\mu^2$ is chosen as $m_t^2$, and the dimensionless variables are
\begin{equation}
  \label{ttW_dimless}
  x 
  \equiv - \frac{s_{45}}{m_t^2}
  \, ,
  \quad 
  y 
  \equiv - \frac{s_{14}}{m_t^2}
  \, ,
  \quad 
  z
  \equiv - \frac{s_{23}}{m_t^2}
  \, ,
  \quad
  k 
  \equiv - \frac{m_W^2}{m_t^2}
  \, .
\end{equation}

For this family, we will only consider the elliptic sector in the right panel of Fig.~\ref{ttW_fig}, and the starting master integrals are chosen as
\begin{equation}
  \label{ttW_basis}
  \begin{aligned} 
    \big\{&
    I_{111111000}, %1
    I_{1111110(-1)0}, %2
    I_{1111110(-2)0} %3
    \big\} 
    \, .
  \end{aligned}
\end{equation}  

The maximal cut of the master integral $I_{111111000}$ is given by
\begin{equation}
  \label{maxcut_ttW}
  {\rm MaxCut}[I_{111111000}]
  = \int_{\cal C} \, u (z_8) \frac{{\rm d} z_8}{\sqrt{(x + y + 1)^2 - 4 k z}}
  \, .
\end{equation}
The twist $u (z_8)$ is defined by
\begin{equation}
  \label{u_ttW}
  u (z_8)
  = [P_4 (z_8)]^{- 1 / 2} \prod_{i = 1}^4 (z_8 - c_i)^{- \beta_i \varepsilon}
  \, ,
\end{equation}
and the branch points can be expressed with the dimensionless variables as
\begin{equation}
  \label{branch_ttW}
  c_1 
  = 0
  \, ,
  \quad
  c_2 
  = \frac{a - b}{(x + y + 1)^2 - 4 k z}
  \, ,
  \quad
  c_3 
  = \frac{a + b}{(x + y + 1)^2 - 4 k z}
  \, ,
  \quad
  c_4 
  = 4
  \, ,
\end{equation}
where 
\begin{subequations}
  \label{ab_ttW}
  \begin{align}
    a
    & = - 2 k^2 z + k [x^2 + x (y + 2 z + 1) + 2 (z + 1) (y - z) - x (x + y + 1) (x + y - z)]
    \, ,
    \\
    b 
    & = 2 \sqrt{- k [- y z (k + x) - k (x + 1) y + (k + 1) z (k - x) + k z^2 + x y (x + y + 1)][- z (k + z + 1) + (x + y) (z + 1)]}
    \, .
  \end{align}
\end{subequations}

As before, the relevant variable changes can be obtained combining Eqs.~\eqref{branch_ttW}, \eqref{ab_ttW}, \eqref{lambda} and \eqref{z2_Legendre}. The canonical basis can then be derived by plugging the variables into Eq.~\eqref{canbasis}. Due to the complicated kinematics in this family, the explicit expressions are rather lengthy and not too illuminating. We leave them in the ancillary files attached to this paper.

\subsection{An elliptic branch of $2$-loop $W$-pair production via light quark-antiquark annihilation}

\label{subsec_t3f}

\begin{figure}
  \centering
\includegraphics[width=15cm]{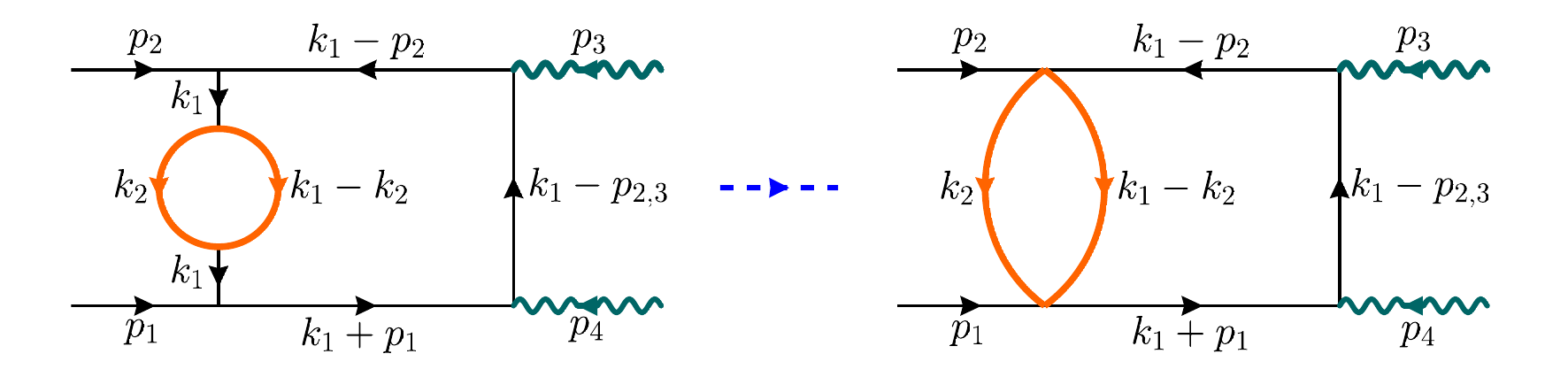}
  \caption{The ${\cal T}_{3F}$ branch of $2$-loop $W$-pair production via light quark-antiquark annihilation~\cite{He:2024iqg} is shown in the left panel. The sector is reducible, and a special sub-sector is shown in the right panel. This sub-sector is the only elliptic sector. In the diagrams, thick orange lines represent top quarks with mass $m_t$ and the curvy lines represent $W$ bosons with mass $m_W$ while all the others are massless.}
  \label{t3f_fig}
\end{figure}

The diagram for the ${\cal T}_{3F}$ branch of $2$-loop $W$-pair production via light quark-antiquark annihilation is shown in Fig.~\ref{t3f_fig}. We follow the notations in~\cite{He:2024iqg}, and the propagators are 
\begin{equation}
  \label{t3f_prop}
  \begin{aligned}
    D_1 
    & = k_1^2 
    \, , 
    \quad &
    D_2 
    & = (k_1 + p_1)^2
    \, , 
    \quad &
    D_3 
    & = (k_1 - p_2)^2 
    \, ,
    \\
    D_4 
    & = k_2^2 - m_t^2 
    \, , 
    \quad &
    D_5 
    & = (k_2 + p_1)^2 -m_t^2
    \, ,
    \quad &
    D_6 
    & = (k_2 - p_2)^2 - m_t^2
    \, , 
    \\
    D_7
    & = (k_1 - k_2)^2 - m_t^2
    \, ,
    \quad &
    D_8
    & = (k_1 - p_2 - p_3)^2
    \, ,
    \quad &
    D_9 
    & = (k_2 - p_2 - p_3)^2
    \, .
  \end{aligned}
\end{equation}
The kinematic variables are defined as
\begin{equation}
  \label{t3f_kin}
  s
  \equiv (p_1 + p_2)^2
  \, ,
  \quad 
  t 
  \equiv (p_2 + p_3)^2
  \, .
\end{equation}
The scale $\mu^2$ is chosen as $m_t^2$, then the dimensionless variables are
\begin{equation}
  \label{t3f_dimless}
  x 
  \equiv - \frac{s}{m_t^2}
  \, ,
  \quad 
  y 
  \equiv - \frac{t}{m_t^2}
  \, ,
  \quad 
  z 
  \equiv - \frac{m_W^2}{m_t^2}
  \, .
\end{equation}

For this family, we are going to derive the full canonical basis without any cuts. The starting master integrals are chosen as
\begin{equation}
  \label{t3f_basis}
  \begin{aligned} 
    \big\{&
    I_{011200110}, %1
    I_{(-1)11100110}, %2
    I_{(-1)11200110}, %3
    I_{111100010}, %4
    I_{(-1)10100110}, %5
    \\ &
    I_{010100110}, %6 
    I_{(-1)11100100}, %7 
    I_{011100100}, %8
    I_{011100010}, %9
    I_{100200010}, %10
    \\ &
    I_{(-1)00100110}, %11
    I_{000100110}, %12
    I_{010100010}, %13
    I_{011100000}, %14
    I_{000200200} %15
    \big\} 
    \, .
  \end{aligned}
\end{equation}
The first $3$ master integrals belong to the elliptic top sector. The remaining master integrals belong to non-elliptic sub-sectors, whose canonical bases can be easily constructed with ${\rm d \log}$-forms.

The maximal cut of the master integral $I_{011200110}$ is given by
\begin{equation}
  \label{maxcut_t3f}
  {\rm MaxCut}[I_{011200110}]
  = \int_{\cal C} \, u (z_1) \frac{1 - 2 \varepsilon}{\sqrt{x (x - 4 z)}} {\rm d} z_1
  \, ,
\end{equation}
where
\begin{equation}
  \label{u_t3f}
  u (z_1)
  = [P_4 (z_1)]^{- 1 / 2} \prod_{i = 1}^4 (z_1 - c_i)^{- \beta_i \varepsilon}
  \, ,
\end{equation}
and the branch points can be expressed with the dimensionless variables as
\begin{equation}
  \label{branch_t3f}
  c_1 
  = 0
  \, ,
  \quad
  c_2 
  = \frac{2 z (y - z) - x y}{x - 4 z} + \frac{z \sqrt{x y + (y - z)^2}}{x - 4 z}
  \, ,
  \quad
  c_3 
  = \frac{2 z (y - z) - x y}{x - 4 z} - \frac{z \sqrt{x y + (y - z)^2}}{x - 4 z}
  \, ,
  \quad
  c_4 
  = 4
  \, .
\end{equation}
After M\"obius transformation, the relevant variables are
\begin{equation}
  \label{variable_t3f}
  \lambda
  = \frac{16 z \sqrt{x y + (y - z)^2}}{8 z \left(\sqrt{x y + (y - z)^2} - y + z\right) + x y (y + 4)}
  \, ,
  \quad
  e_5 
  = \frac{4 (x - 4 z)}{2 z \left(- \sqrt{x y + (y - z)^2} - y + z - 8\right) + x (y + 4)}
  \, .
\end{equation}
Plugging the variables into Eq.~\eqref{canbasis}, we obtained the integrands for the canonical basis under maximal cut.

At this point, it is worth pointing out that there are always ambiguities when promoting the integrands with cuts to Feynman integrals without cuts. The sub-sector dependence of the resulting integrals strongly depends on the promoting procedure. In our case, we simply convert the canonical bases in Eq.~\eqref{canbasis} into linear combinations of the top-sector master integrals in Eq.~\eqref{t3f_basis}, without adding further components from sub-sectors. Therefore, the sub-sector dependence is determined by the choice of the top-sector master integrals. With the choice in Eq.~\eqref{t3f_basis}, we find that the sub-sector dependence is mostly $\varepsilon$-factorized, except for $\phi_{n-1}^{\rm can}$ in Eq.~\eqref{canbasis} that involves the differential of the second kind of elliptic integral.

It has been noticed in many situations that the derivative of the top-sector integral corresponding to $\phi_1^{\rm can}$ leads to a simpler sub-sector dependence. In the current case, this corresponds to the derivative of
\begin{equation}
  \frac{2 \varepsilon^3}{\varpi_1(\lambda)} \sqrt{- 8 x z \big[\sqrt{x y + (y - z)^2} - y + z\big] - x^2 y (y + 1)} \, I_{011200110} \,.
\end{equation}
Note that this is NOT the same as the derivative in \eqref{precan_new}, since the latter lives under the maximal cut with no information about sub-sectors. In other words, while the canonical forms are uniquely fixed under the maximal cut, their extension to full Feynman integrals is not unique. One has the freedom to add sub-sector integrals which vanish under the maximal cut. These choices directly determine the complexity of the off-diagonal blocks. By using the information from the derivative, we can identify specific ``good'' combinations of sub-sector integrals to add when promoting $\phi_{n-1}^{\rm can}$ to Feynman integrals without cuts. We will employ this trick for the examples in Sec.~\ref{subsec_npdb} and Sec.~\ref{subsec_npdb2} as well. When this is done (where the derivative is taken with respect to $x$), we find that there are only 3 entries in the differential equations which are not $\varepsilon$-factorized.

We now simply need to rotate away the $\varepsilon^0$-terms in the $3$ entries to reach a fully $\varepsilon$-factorized (canonical) differential equation. This can be done via simple one-fold integrations. The explicit expressions are rather lengthy, and we do not present them here. The final canonical basis integrals are given in ancillary files.

\subsection{Non-planar double box A}

\label{subsec_npdb}

\subsubsection{Definition and two-variable construction under the next-to-maximal cut}

\label{subsubsec_def_npdb}

\begin{figure}
  \centering
  \includegraphics[width=15cm]{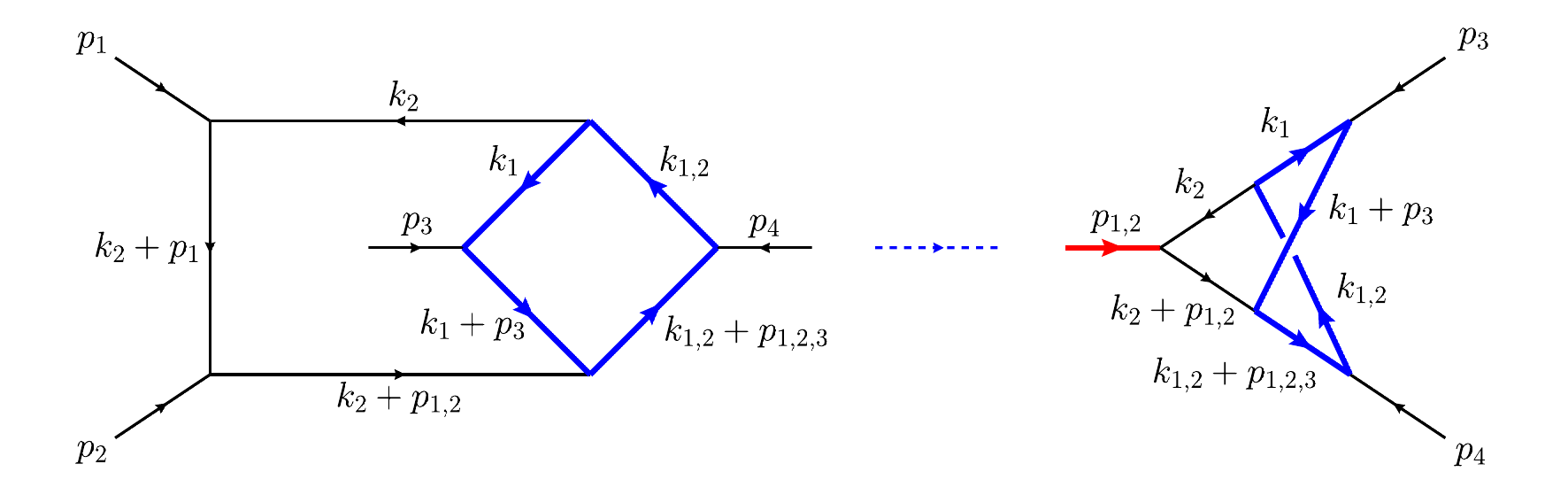}
  \caption{The diagram for the non-planar double box family A is shown in the left panel. The sub-sectors are obtained by pinching the propagators. A special case is the elliptic triangle sector, shown in the right panel. This sub-sector and the top sector are the only sectors that are elliptic, and they share the same elliptic curve. In the diagrams, thick blue lines are massive while all the others are massless, and we use the shorthand $p_{1,2,3}=p_1+p_2+p_3$ and $k_{1,2}=k_1+k_2$ hereafter.}
  \label{npdb_fig}
\end{figure}

The non-planar double box diagram is shown in Fig.~\ref{npdb_fig}. All four external legs are massless and incoming, and it has a massive internal loop. $D_1,\cdots, D_7$ are seven propagators, shown in Fig.~\ref{npdb_fig}, while $D_8$ and $D_9$ are two ISPs, i.e., $v_8 \leq 0,v_9 \leq 0$. They are defined as
\begin{equation} 
  \label{props}
  \begin{aligned}
  D_1 
  & = k_1^2 - \chi m^2 
  \, , 
  \quad &
  D_2 
  & = (k_1 + p_3)^2 - m^2 
  \, , 
  \quad &
  D_3 
  & = (k_1 + k_2 + p_1 + p_2 + p_3)^2 - \chi m^2 
  \, ,
  \\
  D_4 
  & = (k_1 + k_2)^2 - m^2 
  \, , 
  \quad &
  D_5 
  & = k_2^2 
  \, ,
  \quad &
  D_6 
  & = (k_2 + p_1)^2 
  \, , 
  \\
  D_7 
  & = (k_2 + p_1 +p_2)^2 
  \, ,
  \quad &
  D_8 
  & = (k_1 - p_1)^2 - m^2 
  \, , 
  \quad &
  D_9 
  & = (k_2 + p_1 + p_3)^2 
  \, ,
  \end{aligned}
\end{equation}
where we have $\chi = 1$ in this example. The integral family depends on the mass $m$ and two Mandelstam variables associated with the incoming momenta $p_1$, $p_2$ and $p_3$:
\begin{equation}
  \label{npdb_kin}
	s 
  \equiv (p_1 + p_2)^2 
  \, ,
  \quad 
  t 
  \equiv (p_1 + p_3)^2
  \, .
\end{equation}
We have defined the integrals as dimensionless by introducing the scale $\mu^2$. We will choose $\mu^2=s$, such that the integrals are functions of the dimensionless variables
\begin{equation}
  \label{npdb_dimless}
	y_s 
  \equiv 
  - \frac{m ^2}{s}
  \, ,
  \quad 
  y_t 
  \equiv - \frac{t}{s}
  \, .
\end{equation}

We use \texttt{Kira}~\cite{Klappert:2020nbg} to reduce the integrals in this family to a set of master integrals, chosen as
\begin{equation}
  \label{npdb_basis}
  \begin{aligned} 
    \big\{&
    I_{111111100}, %1
    I_{11111110(-1)}, %2
    I_{21111110(-1)}, %3
    I_{11121110(-1)}, %4
    I_{111110100}, %5
    I_{111210100}, %6
    \\ &
    I_{111101100}, %7
    I_{111201100}, %8
    I_{110111100}, %9
    I_{111011100}, %10
    I_{20101110(-1)}, %11
    I_{201011100}, %12
    \\ &
    I_{010211100}, %13
    I_{010111100}, %14
    I_{110201100}, %15
    I_{110101100}, %16
    I_{111102000}, %17
    I_{121101000}, %18
    \\ &
    I_{111101000}, %19
    I_{201101100}, %20
    I_{101101100}, %21
    I_{111010100}, %22
    I_{111100100}, %23
    I_{102010200}, %24
    \\ &
    I_{102010100}, %25
    I_{110200100}, %26
    I_{110201000}, %27
    I_{112001000}, %28
    I_{002020100}, %29
    I_{010202000}, %30
    \\ &
    I_{020201000}, %31
    I_{200100200}, %32
    I_{200200100}, %33
    I_{201002000}, %34
    I_{202001000}, %35
    I_{022000000} %36
    \big\} 
    \, .
  \end{aligned}
\end{equation}
The first 4 master integrals are in the top sector. They are chosen such that their differential equations have relatively simple dependencies on the (canonical) sub-sector integrals. This will be helpful in transforming these dependencies into the canonical form. Note that our construction for the top sector integrals under the maximal cut does not depend on the choice, and different choices are the same the under the maximal cut while lead to different sub-sector dependence. The last two master integrals in the first line of Eq.~\eqref{npdb_basis} belong to the elliptic triangle family, whose canonical bases are well-studied in the literature~\cite{Jiang:2023jmk,Gorges:2023zgv}. For the remaining sub-sectors, we construct the canonical basis integrals with ${\rm d} \log$-forms following~\cite{Chen:2020uyk,Chen:2022lzr}, and the explicit expressions for them are given in App.~\ref{app_basis}. The canonical bases for these non-elliptic sub-sectors were also studied in \cite{Ahmed:2024tsg,Becchetti:2025rrz}.

In this family, we have two sectors sharing the same elliptic curve. Therefore, although a canonical basis for this family has been recently obtained in \cite{Becchetti:2025rrz}, we find it enlightening to study the two elliptic sectors together by applying the next-to-maximal elliptic triangle cut instead of the maximal cut. Under the triangle cut, we have two integration variables remaining. For example, the master integral $I_{111110100}$ in the elliptic triangle sector can be written as
\begin{equation}
  \label{maxcutdb}
  {\rm EllCut} [I_{111110100}] 
   =  \int_{\cal C} u(z_6, z) \, {\rm d} z_6 \wedge {\rm d} z 
  \, ,
\end{equation}
where we have defined $z \equiv z_9 - z_6 + c_5$, and the twist is given by
\begin{equation}
  \label{twist_npdb}
  u(z_6, z) = [P_4 (z)]^{-1/2} [P_2 (z_6, z)]^{-1/2} \prod_{i = 1}^4 (z - c_i )^{- \beta_i \varepsilon} \prod_{j = 6}^7 (z_6 - c_j)^{- \beta_j \varepsilon} 
  \, .
\end{equation}
The branch points ${c_1, c_2, c_3, c_4}$ can be expressed with kinematic variables $y_s$ and $y_t$ as
\begin{equation}
  \label{branch_point_npdb}
  c_1 
  = 0 
  \, , 
  \quad 
  c_2 
  = \frac{1}{2}\left(1 - \sqrt{1 - 16 y_s}\right) 
  \, , 
  \quad
  c_3 
  = \frac{1}{2}\left(1 + \sqrt{1 - 16 y_s}\right) 
  \, , 
  \quad
  c_4 
  = 1 
  \, , 
\end{equation}
while the branch points $c_6$ and $c_7$ depend also on the variable $z$:
\begin{equation}
  \label{branch_point_npdb_2}
  c_{6,7}
  = 2 y_t z -y_t - z \pm 2 \sqrt{y_t (y_t - 1) z (z - 1)}
  \, .
\end{equation}
The polynomial $P_2 (z_6, z)$ is given by
\begin{equation}
  \label{P2_npdb}
  P_2 (z_6, z)
  = (z_6 - c_6) (z_6 - c_7)
  \, .
\end{equation}

Since the variable $z_6$ appears quadratically under the square root in the twist, we can construct a $\mathrm{d} \log$ with respect to it, and arrive at
\begin{equation}
  \label{ellwedgedlog_tri}
  {\rm EllCut} [I_{111110100}]
  = \int_{\cal C} \, \bar{u} (z_6, z) \frac{{\rm d} z_6}{\sqrt{P_2(z_6, z)}} \wedge \frac{{\rm d} z}{\sqrt{P_4(z)}}
  = - \int_{\cal C} \, \bar{u} (z_6, z) {\rm d} \log \left(\frac{1 + \sqrt{\frac{z_6 - c_6}{z_6 - c_7}}}{1 - \sqrt{\frac{z_6 - c_6}{z_6 - c_7}}}\right) \wedge \frac{{\rm d} z}{\sqrt{P_4(z)}} 
  \, ,
\end{equation}
where $\bar{u} (z_6, z)$ is the reduced twist
\begin{equation}
  \label{reduced_twist}
  \bar{u} (z_6, z)
  = \prod_{i = 1}^4 (z - c_i )^{- \beta_i \varepsilon} \prod_{j = 6}^7 (z_6 - c_j)^{- \beta_j \varepsilon} 
  \, .
\end{equation}
Similarly, the top-sector master integral $I_{111111100}$ under the triangle cut can be written as
\begin{align}
  \label{ellwedgedlog_top}
  {\rm EllCut} [I_{111111100}]
  & = \int_{\cal C} \, \bar{u} (z_6, z) \frac{{\rm d} z_6}{z_6 \sqrt{P_2(z_6, z)}} \wedge \frac{{\rm d} z}{\sqrt{P_4(z)}}
  \\ 
  & = \int_{\cal C} \, \bar{u} (z_6, z) {\rm d} \log \left(\frac{1 + \sqrt{\frac{c_7 (z_6 - c_6)}{c_6 (z_6 - c_7)}}}{1 - \sqrt{\frac{c_7 (z_6 - c_6)}{c_6 (z_6 - c_7)}}}\right) \wedge \frac{{\rm d} z}{(z - c_5) \sqrt{P_4(z)}}
  \, ,
\end{align}
where the branch point $c_5=y_t$ for $z$ appears after constructing the $\mathrm{d}\log$ for $z_6$:
\begin{equation}
  P_2(0,z) = c_6 c_7 = (z-c_5)^2 \,.
\end{equation}

From Eqs.~\eqref{ellwedgedlog_tri} and~\eqref{ellwedgedlog_top}, we can see explicitly that they share the same elliptic curve defined by the degree-4 polynomial $P_4 (z)$. Therefore, we may utilize the construction in Sec.~\ref{sec:canbasis} to write down the integrands for $z$, such that the integrands under the next-to-maximal cut have the form ${\rm d} \log \wedge \phi^{\rm can}$.
Applying the M\"obius transformation in Sec.~\ref{sec:canbasis}, we arrive at the following variables:
\begin{equation}
  \label{variable_npdb}
  \lambda 
  = \frac{4 \sqrt{1 - 16 y_s}}{\left(1 + \sqrt{1 - 16 y_s}\right)^2}
  \, , 
  \quad
  e_5
  = \frac{\sqrt{1 - 16 y_s} + 2 y_t - 1}{\left(1 + \sqrt{1 -16 y_s}\right) y_t}
  \, ,
  \quad 
  e_6 
  = \frac{1 - \sqrt{1 - 16 y_s}}{8 y_s}
  \, .
\end{equation}
Note that $\lambda$ is related to the variable $k$ in \cite{Jiang:2023jmk} as
\begin{equation}
  \lambda=1-k^2 \,, \quad k = \frac{1 - \sqrt{1 - 16 y_s}}{1 + \sqrt{1 + 16 y_s}} \,.
\end{equation}

We now just need to plug the relevant variables into Eq.~\eqref{canbasis} to obtain $\phi^{\rm can}$. At this point, we would like to point out an interesting phenomenon. The integrands in Eq.~\eqref{canbasis} involve a complete elliptic integral of the third kind depending on $e_6$, for both the top sector and the elliptic triangle sector. However, from the results of \cite{Jiang:2023jmk}, we have already known that no such integral is needed for the triangle sector. This hints at relations between complete elliptic integrals of the first kind and the third kind.\footnote{One may derive similar relations from the sunrise families. In both cases, we find that such relations are associated with the fact that the integrands of the first kind and the third kind can be combined into a ${\rm d} \log$-form.}
In the current case, we have
\begin{equation}
  \label{special_relation_npdb}
  \vartheta_0 (e_6, \lambda) 
  = - \frac{\varpi_0 (\lambda)}{1 + \sqrt{1 - 16 y_s}} 
  \, ,
  \quad
  \vartheta_1 (e_6, \lambda)
  = - \frac{\varpi_1 (\lambda)}{1 + \sqrt{1 - 16 y_s}} - 1
  \, .
\end{equation}
Inserting the explicit forms of $\varpi_i$ and $\vartheta_i$, we arrive at the relations
\begin{equation}
  \label{special_relation_sim_npdb}
  \Pi (1 - k, 1 - k^2)
  = \frac{(1 + k) K (1 - k^2)}{2 k}
  \, ,
  \quad
  \Pi (k, k^2) 
  = \frac{1}{2} K (k^2) + \frac{\pi}{4 (1 - k)}
  \, .
\end{equation}

The two-variable integrands encode more information than the univariate integrands for each of the two elliptic sectors. Therefore, it is interesting to investigate the differential equations of the constructed basis, especially the mixing between these two sectors. We find that the differential equations are $\varepsilon$-factorized within each of the two sectors, as expected. However, the mixing terms are not completed factorized. The problem is again associated with $\phi_{n-1}^{\rm can}$ in the top-sector. We employ the derivative trick as in Sec.~\ref{subsec_t3f} (this time with respect to $y_s$) to find suitable combinations of sub-sector integrals to be added. When this is done, we find that all non-$\varepsilon$-factorized terms are removed. This is of course an accident, since we have seen in Sec.~\ref{subsec_t3f} that the derivative trick is not always able to get rid of all $\varepsilon^0$ terms. We will see similar behaviors in the non-planar double-box family with a different mass configuration in Sec.~\ref{subsec_npdb2}.

\subsubsection{Mixing with non-elliptic sectors}

\label{subsubsec_sim_npdb}

With the two elliptic sectors sorted out, we now go on to derive the full canonical basis without cuts. For that we only need to clean the mixing between the elliptic top-sector and the non-elliptic sub-sectors. We find that such mixing in this family is considerably more complicated than the previous examples, even after applying the derivative trick. Part of the reason is that elliptic integrals appear at multiple places in the canonical integrands \eqref{canbasis}. Therefore, we find it useful to study the sub-sector dependence of the pre-canonical basis \eqref{pre_basis_general} (promoted to Feynman integrals without cuts), before rotating it to the canonical one. In particular, if we manage to make the sub-sector dependence $\varepsilon$-factorized for the pre-canonical integrals of the first and the third kind, the same structure will be preserved when rotating to the canonical ones. Afterwards, we only need to further deal with the canonical integral of the second kind.

The coefficient matrix $\bm{A}$ in the differential equations for the pre-canonical basis $\vec{I}^{\text{pre}}$ within the top-sector has the following structure:
\begin{equation}
  \label{top_pre_de}
  \left(\begin{array}{cccc}
    \cellcolor{Purple} {\bm A}_{1,1} & \cellcolor{cyan} {\bm A}_{1,2} & \cellcolor{cyan} {\bm A}_{1,3} & \cellcolor{red} {\bm A}_{1,4} 
    \\
    \cellcolor{Purple} {\bm A}_{2,1} & \cellcolor{cyan} {\bm A}_{2,2} & \cellcolor{cyan} {\bm A}_{2,3} & \cellcolor{red} {\bm A}_{2,4} 
    \\
    \cellcolor{Purple} {\bm A}_{3,1} & \cellcolor{cyan} {\bm A}_{3,2} & \cellcolor{cyan} {\bm A}_{3,3} & \cellcolor{red} {\bm A}_{3,4} 
    \\
    \cellcolor{Purple} {\bm A}_{4,1} & \cellcolor{cyan} {\bm A}_{4,2} & \cellcolor{cyan} {\bm A}_{4,3} & \cellcolor{Purple} {\bm A}_{4,4} 
  \end{array}\right)
    \, ,
\end{equation}
where the cyan cells are already $\varepsilon$-factorized, the purple cells are linear in $\varepsilon$, and the red cells contain only $\varepsilon^0$-terms. We would like to add sub-sector components to $\vec{I}^{\text{pre}}$ such that the above structure is not spoiled, and we can then rotate it to the canonical basis using the transformation matrix $\mathcal{T}_{\text{can}}$ derived in Sec.~\ref{sec:canbasis}.

We first deal with the $\varepsilon^0$-terms in the sub-sector dependence of $I_4^{\text{pre}}$, which only appear in the matrix element ${\bm A}_{4,8}$. We denote the canonical basis for the non-elliptic sub-sectors as $M_j$ for $j=7,\cdots,36$. Introducing two coefficient functions $f_{1,8}$ and $f_{4,8}$, we apply the following transformation to $I_1^{\text{pre}}$ and $I_4^{\text{pre}}$:
\begin{equation}
  \label{ansatzpre42mpl_npdb}
  I_1^{\text{pre}} 
  \to I_1^{\text{pre}} + f_{1, 8} (y_s, y_t) M_8
  \, ,
  \quad
  I_4^{\text{pre}}
  \to I_4^{\text{pre}} + f_{4, 8} (y_s, y_t) M_8
  \, .
\end{equation}
Requiring that the $\varepsilon^0$-terms vanish in the transformed differential equations, we can find the solution for the two functions:
\begin{equation}
  \label{ansatzpre42mpl_npdb_sol}
  f_{1, 8} (y_s, y_t)
  = - \frac{4 (1 + \sqrt{1 - 16 y_s}) (2 y_t - 1)}{\sqrt{y_s (y_t - 1) y_t}}
  \, ,
  \quad
  f_{4, 8} (y_s, y_t)
  = - \frac{8 \sqrt{1 - 16 y_s} (2 y_t - 1)}{(1 + \sqrt{1 - 16 y_s}) \sqrt{y_s (y_t - 1) y_t}}
  \, .
\end{equation}

After the above transformation, the matrix elements ${\bm A}_{1,j}$ for $j = 7, \cdots, 36$ are all $\varepsilon$-factorized, while ${\bm A}_{4,j}$ are in the form $a^{(1)} \varepsilon + a^{(2)} \varepsilon^2$. We now turn to the $\varepsilon^0$-terms in the sub-sector dependence of $I_{2}^{\text{pre}}$ and $I_{3}^{\text{pre}}$. According to \eqref{top_pre_de}, all the matrix elements $A_{i,2/3}$ are already $\varepsilon$-factorized. Therefore, we can apply the following transformations to $I_{2/3}^{\text{pre}}$:
\begin{equation}
  \label{ansatzpre232mpl_npdb}
  I_{2/3}^{\text{pre}}
  \to I_{2/3}^{\text{pre}} + f_{2/3,j} (y_s, y_t) M_j
  \, ,
\end{equation}
without altering the $\varepsilon$-factorized structure for $I_1^{\text{pre}}$ and $I_4^{\text{pre}}$. The functions $f_{2/3,j} (y_s, y_t)$ can be solved by requiring the vanishing of $\varepsilon^0$-terms. Their expressions are rather lengthy, and we choose to not present them here.

With the above preparation, we can now apply the transformation ${\cal T}_{\rm can}$ to convert $\vec{I}^{\text{pre}}$ into $\vec{I}^{\text{can}}$. From the structure of ${\cal T}_{\rm can}$, it is easy to see that the sub-sector dependence of $I^{\text{can}}_{1/2/3}$ is still $\varepsilon$-factorized. On the other hand, the $\varepsilon^0$-terms re-appear in the differential equations of $I^{\text{can}}_{4}$, which is now in the linear form $a^{(0)} + a^{(1)} \varepsilon$. These $\varepsilon^0$-terms can be removed by one-fold integrations as before. The final canonical basis for Feynman integrals without cuts is given in ancillary files.

\subsection{Non-planar double box B}

\label{subsec_npdb2}

\begin{figure}
  \centering
  \includegraphics[width=15cm]{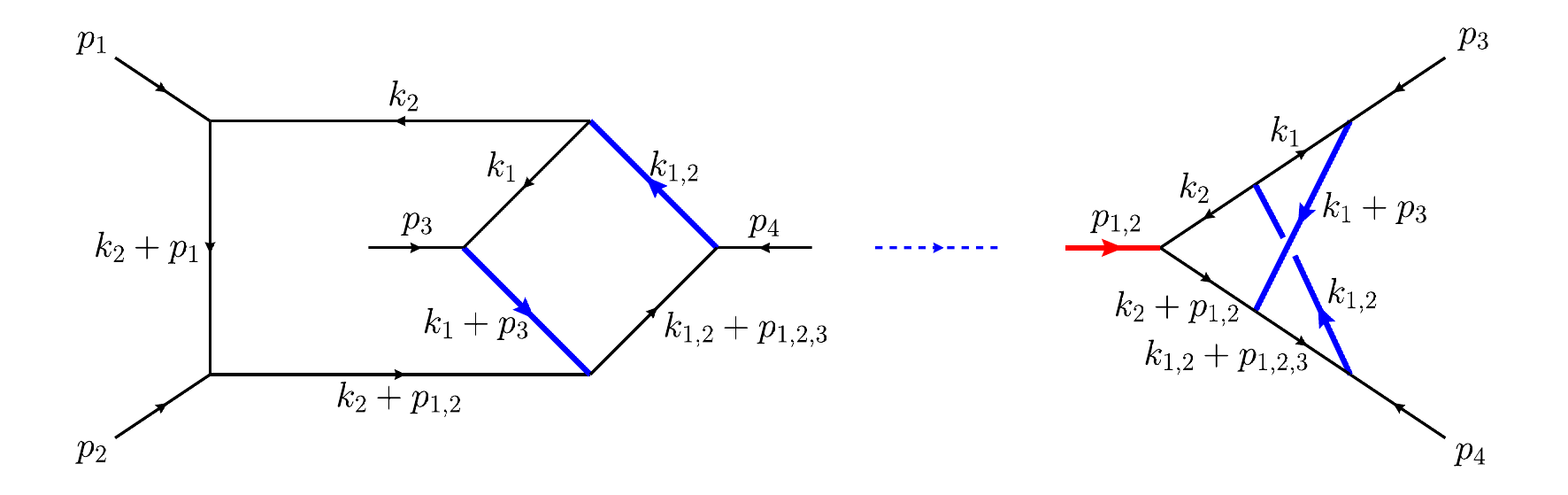}
  \caption{The diagram for the non-planar double box family B is shown above. It is similar to Fig.~\ref{npdb_fig}, with two internal lines becoming massless.}
  \label{npdb2_fig}
\end{figure}

Our final example is the non-planar double box B family shown in Fig.~\ref{npdb2_fig}. It shares a lot of similarities with the family in Sec.~\ref{subsec_npdb}, and the canonical basis is studied in~\cite{Schwanemann:2024kbg}. The kinematics is the same as Eqs.~\eqref{npdb_kin} and~\eqref{npdb_dimless}. The propagators are also the same as Eq.~\eqref{props} except that $\chi = 0$, so that $D_1$ and $D_3$ are massless. For this family, we will only consider the top sector and the elliptic triangle sub-sector\footnote{Note that there is actually another elliptic curve associated with a different sub-sector, which we do not consider here.}, which share the same elliptic curve. The starting master integrals are chosen as
\begin{equation}
  \label{npdb2_basis}
  \begin{aligned} 
    \big\{&
    I_{21111110(-1)}, %1
    I_{11121110(-1)}, %2
    I_{11111110(-1)}, %3
    I_{111111100}, %4
    I_{211110100}, %5
    I_{111210100}, %6
    I_{111110100} %7
    \big\} 
    \, .
  \end{aligned}
\end{equation}
The first $4$ integrals belong to the top sector, while the other $3$ belong to the triangle sub-sector. We have one more master integral in the triangle sector than the case in Sec.~\ref{subsec_npdb} because the diagram has fewer symmetry relations.

As the case in Sec.~\ref{subsec_npdb}, we perform two-variable constructions in the form ${\rm d} \log \wedge \phi^{\rm can}$ under the elliptic triangle cut, where the mixing between the two sectors is taken into account. The elliptic cut of $I_{111111100}$, $I_{111110100}$ and the reduced twist shares the same form as Eqs.~\eqref{ellwedgedlog_top}, \eqref{ellwedgedlog_tri} and \eqref{reduced_twist}, respectively. Defining $z = - z_9 + z_6 + c_5$, we can express the branch points ${c_1, c_2, c_3, c_4}$ for this family as
\begin{equation}
  \label{branch_point_ewdb}
  c_1
  = 0 
  \, , 
  \quad 
  c_2
  = \frac{(1 - x)^2}{4}
  \, ,
  \quad
  c_3
  = \frac{(1 + x)^2}{4}
  \, , 
  \quad
  c_4
  = 1 
  \, ,
\end{equation}
while $c_5$, $c_6$ and $c_7$ are given by
\begin{equation}
  \label{branch_point_ewdb_2}
  c_5 
  = y 
  \, ,
  \quad
  c_{6,7}
  = y (2 z - 1) - z \pm 2 \sqrt{y (y - 1) z (z - 1)}
  \, ,
\end{equation}
where we have introduced the variables $x$ and $y$ to simplify the expressions:
\begin{equation}
  \label{xy}
  x 
  \equiv \sqrt{1 - 8 y_s}
  \, ,
  \quad
  y
  \equiv 1 - y_t
  \, .
\end{equation}

After the M\"obius transformation, the relevant variables are
\begin{equation}
  \label{variable_ewdb}
  \lambda 
  = \frac{16 x}{(3 - x) (1 + x)^3}
  \, , 
  \quad
  e_5
  = \frac{4 y - x^2 + 2 x - 1}{y (3 - x) (1 + x)}
  \, ,
  \quad 
  e_6 
  = \frac{4}{(3 - x) (1 + x)}
  \, .
\end{equation}
Inserting the variables into Eq.~\eqref{canbasis}, we readily obtain the integrands for the canonical basis. As before, we may find special relations similar to Eqs.~\eqref{special_relation_npdb} and~\eqref{special_relation_sim_npdb}, which are
\begin{equation}
  \label{special_relation_npdb2}
  \vartheta_0 (e_6, \lambda)
  = - \frac{(x^2 + 3) \varpi_0 (\lambda)}{3 (1 + x) \sqrt{(3 - x) (1 + x)}}
  \, ,
  \quad
  \vartheta_1 (e_6, \lambda)
  = - \frac{(x^2 + 3) \varpi_1 (\lambda)}{3 (1 + x) \sqrt{(3 - x) (1 + x)}} - \frac{4}{3}
  \, .
\end{equation}
The explicit forms are 
\begin{align}
  \label{special_relation_sim_npdb2}
  \Pi \left(\frac{4 x}{(1 + x)^2}, \frac{16 x}{(3 - x) (1 + x)^3}\right)
  & = \frac{x^2 + 3}{3 (1 - x)^2} K \left(\frac{16 x}{(3 - x) (1 + x)^3}\right)
  \, ,
  \\
  \Pi \left(\frac{(1 - x) (3 + x)}{(3 - x) (1 + x)}, 1 - \frac{16 x}{(3 - x) (1 + x)^3}\right)
  & = \frac{3 - x}{6} K \left(1 - \frac{16 x}{(3 - x) (1 + x)^3}\right) + \frac{\pi (3 - x) (1 + x)^2}{12 x \sqrt{(3 - x) (1 + x)}}
  \, .
\end{align}

Applying the derivative trick as in earlier examples, we find that there is still one entry $\tilde{{\bm A}}_{4,5}$ in the differential equations that are not $\varepsilon$-factorized. This can be removed by a simple transformation:
\begin{equation}
  \label{restrans_npdb2}
  I_4^{\rm can}
  \to I_4^{\rm can} - \frac{2 x^2 \varpi_1 (\lambda)^2}{3 (x - 3) (x + 1)^3} I_5^{\rm can}
  \, .
\end{equation}
Alternatively, we can also perform the transformation at the level of the pre-canonical basis:
\begin{equation}
  \label{restrans_npdb2_pre}
  I_4^{\rm pre}
  \to I_4^{\rm pre} - \frac{16 x^2 \varepsilon}{3 (x - 3) (x + 1)^3} I_5^{\rm pre}
  \, ,
\end{equation}
before acting the matrix $\mathcal{T}^{\rm can}$. Both methods lead to a canonical basis within the two sectors.

\section{Conclusion and outlook}

\label{sec_conclusion}

In this paper, we present an integrand-level canonical basis Eqs.~\eqref{canbasis} for elliptic Feynman integral families involving multiple kinematic scales, which exhibit a univariate integral in the Baikov representation under maximal cut. The applications to specific integral families merely boil down to plug the information of branch points into Eqs.~\eqref{z2_Legendre} and Eqs.~\eqref{canbasis}. Furthermore, the canonical bases exhibit favorable behaviors around chosen degenerate points, where they asymptotically take the ${\rm d} \log$-forms. This helps to obtain simple boundary conditions for the differential equations. Around the degenerate point, only the dependence of the basis integrals on the marked points becomes significant, allowing the integrals to be expressed in terms of MPLs in this context. 

Our derivation of the canonical basis is to perform the construction using the Legendre normal form of elliptic curves, where the geometric information is explicit, and then apply a M\"obius transformation to map it back to the original integral family, making the results applicable for generic families. The pre-canonical basis integrals under maximal cut are expressed as linear combinations of Abelian differentials without elliptic integrals in the combination coefficients. This feature provides simplification when performing the $\varepsilon$-factorization for the dependence on sub-sectors.  We emphasize again that, while our construction of the pre-canonical bases in the top-sectors and the transformation to $\varepsilon$-factorized (canonical) ones are equivalent to the methods in~\cite{vonManteuffel:2017hms,Adams:2018bsn,Frellesvig:2021hkr,Giroux:2022wav,Dlapa:2022wdu,Pogel:2022vat,Gorges:2023zgv}, these only need to be done once and for all, and the universal results have already been summarized in Eqs.~\eqref{canbasis}.

To demonstrate our method, we apply it to five integral families and obtain canonical basis with success, where two of them are new results in this paper. One of them is under the maximal cut in Sec.~\ref{subsec_ttW}, and the other is for the full integral family without any cuts in Sec.~\ref{subsec_t3f}. Our method can be applied to integral families where multiple sectors involve the same elliptic curve. In such cases, the construction can be done sector by sector. However, we have found that it is helpful to perform the construction for these sectors together by loosening the maximal cut. This has been demonstrated for the non-planar double box families in Sec.~\ref{subsec_npdb}, Sec.~\ref{subsec_npdb2}, where both the top sector and the triangle sub-sector are elliptic.

Our idea of using the Legendre normal form combined with a M\"obius transformation to derive a canonical basis is quite generic. It is largely independent of the particular choice of the pre-canonical bases and the canonicalization transformations. The canonicalization transformation described in the main text is equivalent to the one from~\cite{Gorges:2023zgv}. We have explored several other possibilities which are described in the appendices. In App.~\ref{app_period_eps}, we show that our pre-canonical basis can be transformed to an $\varepsilon$-factorized one by converting the period matrix into a constant matrix without $\varepsilon$-dependent transformation~\cite{vonManteuffel:2017hms,Frellesvig:2021hkr}. Despite a simpler $\varepsilon$-dependence, this method introduces numerous elliptic integrals into the transformation, which, in turn, appear in the $\varepsilon$-factorized differential equations. In addition, this approach inherently leads to the transformation with logarithmic behaviors near degenerate points due to the complementary arguments $\lambda$ and $1 - \lambda$ in elliptic integrals of the first and second kinds. Relevant discussions are included in~\cite{Frellesvig:2023iwr}. In contrast, the approach present in the main text avoids these issues, and the resulting canonical bases satisfy most of the nice properties advocated in~\cite{Dlapa:2022wdu,Frellesvig:2023iwr,Duhr:2024uid}.

Moreover, since the proposed workflow is not inherently limited to elliptic cases, it is expected to extend to more complex geometries, e.g., hyperelliptic cases, as explored in the recent work~\cite{Duhr:2024uid}. For example, using the canonical basis integrals constructed for the four-parameter Lauricella functions, one may derive the canonical basis for the equal-mass non-planar double (crossed) box family, at least under the maximal cut. One can also try constructing a canonical basis in the Rosenhain normal form family with kinematic marked points, which can be potentially applied for generic hyperelliptic Feynman integrals.

It remains possible to extend our two-variable construction in Sec.~\ref{subsec_npdb} and Sec.~\ref{subsec_npdb2}. One may construct the full integrand without any cut, combining the elliptic-sector construction in this work and the ${\rm d} \log$-form construction in \cite{Chen:2020uyk,Chen:2022lzr}. It is also interesting to explore the application of our method to cases where multiple elliptic curves are involved. In this case, there can be non-trivial interplays among the Abelian differentials in different elliptic sectors, and it is interesting to see how these can be handled within our approach. Another complication arises in the boundary conditions since the degenerate points of the different elliptic curves may not coincide. It remains to be seen whether one may utilize the degenerate information separately, and connect them via the differential equations. We leave these for future investigations.

\section*{Acknowledgements}

We would like to thank Ming Lian for collaborations in the early stage of this work.
This work was supported in part by the National Natural Science Foundation of China under Grant No. 12505094, 12375097, 12535003, 12547104, Science Foundation of China University of Petroleum, Beijing (No.2462025YJRC019), and the Fundamental Research Funds for the Central Universities.

\begin{appendix}

\section{Construction of $\varepsilon$-factorized basis with constant period matrix}

\label{app_period_eps}

In Sec.~\ref{subsec_canonical}, we apply the method in~\cite{Pogel:2022vat} to construct the canonical basis. However, it is not the only method to find an $\varepsilon$-factorized basis. With a pre-canonical basis whose connection matrix in a special linear form, we can also try the strategy in~\cite{vonManteuffel:2017hms,Frellesvig:2021hkr}. 

The linear form as Eq.~\eqref{special_linear} is an essential step towards deriving an $\varepsilon$-factorized basis, which will be elaborated soon. In particular, $\bm{A}^{(0)}$ is closely related to the geometry behind the integral family and plays an important role in the later derivation. It is a good place here to put some remarks about it. For this purpose, we take $\varepsilon=0$ in this discussion. With a bit of abuse of notation, we are interested in the following equation:
\begin{equation}
	\label{eq:A0equ}
    {\rm d} \vec{I}^{{\rm pre}, (0)} = {\bm{A}^{(0)}} \, \vec{I}^{{\rm pre}, (0)}
    \,.
\end{equation}
Then based on the general structure given by Eq.~\eqref{lambda_de_0} and \eqref{ei_de_0}, we observe that
\begin{equation}
	\label{eq:A0equelliptic}
	\begin{aligned}
		{\rm d} 
    \begin{pmatrix}
			I_1^{{\rm pre}, (0)}
      \\
			I_{n - 1}^{{\rm pre}, (0)}
		\end{pmatrix} 
    = 
    \begin{pmatrix}
			\frac{{\rm d}\lambda}{2 (1 - \lambda)} & - \frac{{\rm d}\lambda}{2 \lambda (1 - \lambda)} 
      \\
      \frac{{\rm d}\lambda}{2 (1 - \lambda)} & - \frac{{\rm d}\lambda}{2 (1 - \lambda)}
		\end{pmatrix}
    \begin{pmatrix}
			I_1^{{\rm pre}, (0)}
      \\
			I_{n - 1}^{{\rm pre}, (0)}
		\end{pmatrix}
	\end{aligned}
  \, ,
\end{equation}
which equivalently defines the elliptic curve while 
\begin{equation}
	\label{eq:A0equelliptici}
	\begin{aligned}
		{\rm d} I_{i-3}^{{\rm pre}, (0)} 
    = - \left[\frac{e_i(e_i-1)}{(1-\lambda)}\frac{{\rm d}\lambda}{2\sqrt{P_L(e_i)}}+\frac{e_i {\rm d}e_i}{2\sqrt{P_L(e_i)}}\right]I_1^{{\rm pre}, (0)} + \left[\frac{e_i(e_i-1)}{\lambda(1-\lambda)}\frac{{\rm d}\lambda}{2\sqrt{P_L(e_i)}}+\frac{{\rm d}e_i}{2\sqrt{P_L(e_i)}}\right]I_{n-1}^{{\rm pre}, (0)},
	\end{aligned}
\end{equation}
with $5 \leq i \leq n + 1$, which is related to the marked points information in that curve. Denoting
\begin{equation}
	E 
  = E (\sin^{- 1} a_i , \lambda)
  \, ,
  \quad 
  F 
  = F (\sin^{- 1} a_i , \lambda) 
  \, ,
  \quad 
  a_i 
  = \sqrt{e_i / \lambda}
\end{equation}
for brevity, then the right-hand side of the above can be written as a closed form:
\begin{equation}
   \label{eq:A0equellipticicont}
	\begin{aligned}
		{\rm d} I_{i-3}^{{\rm pre}, (0)} &= - \left[{\rm d}(F-E)-E\frac{{\rm d}\lambda}{2(1-\lambda)}\right]I_1^{{\rm pre}, (0)} + \left[{\rm d} F+\left(\frac{F}{2\lambda}-\frac{E}{2\lambda(1-\lambda)}\right){\rm d}\lambda\right]I_{n-1}^{{\rm pre}, (0)}\\
		&= - {\rm d}\left[(F-E)I_1^{{\rm pre}, (0)} - F I_{n-1}^{{\rm pre}, (0)} \right],
	\end{aligned}
\end{equation}
where the second equality in the above follows with the help of the elliptic information \eqref{eq:A0equelliptic}. This property motivates us to refine the master integrals $I_{i-3}^{\rm pre}$ for $5 \leq i \leq n + 1$, which are directly related to the marked points as 
\begin{equation}
  \label{eq:T1}
  I_{i - 3}^{\rm pre} 
  \to \big(\,\mathcal{T}_1 \, \vec{I}^{\rm pre}\,\big)_{i - 3} 
  = I_{i - 3}^{\rm pre} + \left[ F (\sin^{- 1} a_i , \lambda) - E (\sin^{- 1} a_i , \lambda) \right] I_1^{\rm pre} - F (\sin^{- 1} a_i , \lambda) I_{n - 1}^{\rm pre}
  \, .
\end{equation}
Then, their differential equations will be automatically $\varepsilon$-factorized.

Now, the remaining task is to further transform $I_1^{\rm pre}$ and $I_{n - 1}^{\rm pre}$ to remove the corresponding $2 \times 2$ block in $\bm{A}^{(0)}$~\eqref{lambda_de_0}, where the transformation only depends on $\lambda$. To do that, we only need to consider the corresponding $2 \times 2$ block of the period matrix for $I_1^{\rm pre}$ and $I_{n - 1}^{\rm pre}$, which is
\begin{align}
  \label{period_pre}
  {\cal P}
  & =
  \begin{pmatrix}
    2 K (\lambda) & - 2 i K (1 - \lambda)
    \\
    2 K (\lambda) - 2 E (\lambda)  & - 2 i E (1 - \lambda)
  \end{pmatrix}
  \, .
\end{align}
We can see the matrix elements involve linear combinations of the complete elliptic integrals of the first and second kinds. They can be rotated to constants with the help of the well-known \textit{Legendre relation}. A possible rotation matrix is given by
\begin{equation}
  \label{T2}
  {\cal T}_2
  = 
  \begin{pmatrix}
    E (1 - \lambda) & 0 & \cdots & 0 & - K (1 - \lambda)
    \\
    0 & 1 & \cdots & 0 & 0
    \\
    \vdots & \vdots & \ddots & \vdots & \vdots
    \\
    0 & 0 & \cdots & 1 & 0
    \\
    E (\lambda) - K (\lambda) & 0 & \cdots & 0 & K (\lambda)
  \end{pmatrix}
  \, .
\end{equation}
After the rotation, we can easily see that the period matrix becomes a constant matrix.
It is easy to check that the differential equations within the top sector are now $\varepsilon$-factorized after the transformations. The full transformation matrix from the pre-canonical basis to the $\varepsilon$-factorized basis is given by
\begin{equation}
  \label{T}
  {\cal T} 
  = {\cal T}_2 \, {\cal T}_1 = 
  \begin{pmatrix}
    E (1 - \lambda) & 0 & \cdots & 0 & - K (1 - \lambda)
    \\
    F (\sin^{- 1} a_5 , \lambda) - E (\sin^{- 1} a_5 , \lambda) & 1 & \cdots & 0 & - F (\sin^{- 1} a_5 , \lambda) 
    \\
    \vdots & \vdots & \ddots & \vdots & \vdots
    \\
    F (\sin^{- 1} a_\infty , \lambda) - E (\sin^{- 1} a_\infty , \lambda) & 0 & \cdots & 1 & - F (\sin^{- 1} a_\infty , \lambda) 
    \\
    E (\lambda) - K (\lambda) & 0 & \cdots & 0 & K (\lambda)
  \end{pmatrix}
  \, .
\end{equation}
It is straightforward to combine the transformation above with the pre-canonical basis in Eq.~\eqref{pre_basis_general} together with the M\"obius transformation in Eq.~\eqref{z2_Legendre}, to obtain the corresponding $\varepsilon$-factorized basis at the integrand level. This supports the idea that our workflow can be combined with any algorithms to obtain integrand-level results, and then applied to generic integral families.

We've briefly discussed the asymptotic behaviors of the pre-canonical basis in the degenerate limit $c_1 \to c_2$ and thus $\lambda \to 1$. It's natural to consider the asymptotic behaviors of the $\varepsilon$-factorized basis as well. It suffices to study the asymptotic behaviors of the elliptic integrals appearing in the transformation matrices ${\cal T}_1$ and ${\cal T}_2$. We have the following:
\begin{equation}
  \label{elliptic_asymptotic}
  \begin{aligned}
  	&\lim_{\lambda \to 1} K (1 - \lambda) 
  = \lim_{\lambda \to 1} E (1 - \lambda) 
  = \frac{\pi}{2}
  \, , 
  \\
  &\lim_{\lambda \to 1} E (\lambda) 
  = 1
  \, , 
  \qquad 
  \lim_{\lambda \to 1} K (\lambda) 
  = \infty
  \, , 
  \\
  &\lim_{c_1 \to c_2} E (\sin^{- 1} a_i , \lambda)
  = \lim_{\lambda \to 1} E (\lambda) 
  = 1 
  \, , 
  \\
  &\lim_{c_1 \to c_2} F (\sin^{- 1} a_i, \lambda)
  = \lim_{\lambda \to 1} K (\lambda)
  = \infty 
  \, .
  \end{aligned}
\end{equation}
Note that $K (\lambda)$ and $F (\sin^{- 1}a_i, \lambda)$ are singular as $c_1 \to c_2$, and they are logarithmically singular in this limit. These singularities do not cause problems for the basis integrals, since they cancel due to the fact that $\phi_1^{\rm pre} = \phi_2^{\rm pre}$ when $c_1 \to c_2$, see Eq.~\eqref{phi12deglim}. This cancellation also implies that the first $\varepsilon$-factorized basis integral vanishes in this limit. Similar to the pre-canonical basis, other $\varepsilon$-factorized basis integrals are ${\rm d} \log$-forms asymptotically in the neighborhood of the degenerate points $c_1 = c_2$, as desired. 

From the demonstration above, we can see the construction in Sec.~\ref{subsec_canonical} is essentially different from the procedure here from two related perspectives. On the one hand, we introduce $\varepsilon$-dependence in the fiber transformations in the former construction, while the construction here does not. On the other hand, there are only complete elliptic integrals with a unique modulus arguments in the former case, while here we introduce incomplete elliptic integrals and both of complementary moduli arguments are needed. As a consequence, the construction here will introduce more elliptic integrals and undesirable asymptotic properties for the transformation. More relevant discussions can also be found in~\cite{Frellesvig:2023iwr}.

\section{Canonical basis with Jacobi normal form}

\label{app_jacobi}

We construct the pre-canonical basis and the corresponding canonical basis with the elliptic curve in Legendre normal form in Sec.~\ref{sec_construction}. From there, we know the explicit dependence on shape of the curve encoded in the parameter $\lambda$ makes our construction simple. There is another special normal form named Jacobi normal form which shares similar feature with Legendre normal form. Thus, a natural question is why not try constructing the basis in Jacobi form. In this appendix, we will demonstrate the construction of pre-canonical basis and canonical basis in Jacobi normal form. 

The elliptic curve in Jacobi normal form is given by
\begin{equation}
  \label{jacobi}
  v^2 
  \equiv P_J (t)
  \equiv k^2 \,  P_{4J} (t)
  = (1 - t^2) (1 - k^2 t^2)
  \, .
\end{equation}
The twist is
\begin{equation}
  \label{twist_jacobi}
  u_J (t)
  =
  [P_{4J} (t)]^{- 1 / 2} \prod_{i = 1}^{n} (t - a_i)^{- \beta_i \varepsilon}
  \, ,
\end{equation}
with $a_1 = - 1 / k$, $a_2 = - 1$, $a_3 = 1$ and $a_4 = 1 / k$. We also denote 
\begin{equation}
  \beta_{n + 1}
  \equiv
  \beta_\infty
  \equiv
  - \sum_{i = 1}^{n} \beta_i
  \, ,
\end{equation}
and
\begin{equation}
  \phi 
  = \frac{\varphi}{\sqrt{P_{4J} (t)}}
  \, ,
\end{equation}
for convenience.

Since the idea is basically the same as the one in the main text, we will keep the contents concise.

\subsection{The pre-canonical basis}

\label{app_pre}

Similar to the strategy in Sec.~\ref{subsec_construction}, we consider pre-canonical basis integrals as the elliptic integrals of the three kinds in the Jacobi form
\begin{subequations}
  \label{pre_basis_jacobi}
  \begin{align}
    \phi_1^{\rm pre}
    & = \frac{{\rm d} t}{\sqrt{P_J (t)}} 
    = {\rm d} K (\sin^{- 1} t, k^2)
    \, , 
    \\
    \phi_{i - 3}^{\rm pre}
    & = \frac{1}{t - a_i} \frac{\sqrt{P_J (a_i)}}{\sqrt{P_J (t)}} {\rm d} t 
    = \frac{1}{2} {\rm d} \log \left(\frac{1 + \sqrt{\frac{(1 - t^2) (1 - k^2 a_i^2)}{(1 - a_i^2) (1 - k^2 t^2)}}}{1 - \sqrt{\frac{(1 - t^2) (1 - k^2 a_i^2)}{(1 - a_i^2) (1 - k^2 t^2)}}}\right) - \frac{\sqrt{P_J (a_i)}}{a_i} {\rm d} \Pi (1/a_i^2, \sin^{- 1} t, k^2)
    \, ,
    \\
    \phi_{n - 2}^{\rm pre} 
    & = \frac{t}{\sqrt{P_{4J} (t)}} {\rm d} t 
    = \frac{1}{2} {\rm d} \log \left(\frac{1 + k \sqrt{\frac{1 - t^2}{1 - k^2 t^2}}}{1 - k \sqrt{\frac{1 - t^2}{1 - k^2 t^2}}}\right)
    \, ,
    \\
    \phi_{n - 1}^{\rm pre}
    & = \frac{1 - k^2 t^2}{\sqrt{P_J (t)}} {\rm d} t 
    = {\rm d} E (\sin^{- 1} t, k^2)
    \, , 
  \end{align}
\end{subequations}
where $i = 5, \cdots, n$ and the next to last basis differential $\phi_{n - 2}^{\rm pre}$ is a ${\rm d} \log$-form. Then we can find that the differential equation is in a special linear form. The $\varepsilon^0$-part of the connection matrix with respect to $k$ is
\begin{equation}
  \label{de_pre_jacobi_k}
  \begin{pmatrix}
    - \frac{1}{k} & 0 & \cdots & 0 & \frac{1}{k (1 - k^2)} 
    \\
    0 & 0 & \cdots & 0 & - \frac{a_5 (a_5^2 - 1) k}{(1 - k^2) \sqrt{P_J (a_5)}}
    \\
    \vdots & \vdots & \ddots & \vdots & \vdots
    \\
    0 & 0 & \cdots & 0 & - \frac{a_n (a_n^2 - 1) k}{(1 - k^2) \sqrt{P_J (a_n)}}
    \\
    0 & 0 & \cdots & 0 & 0
    \\
    - \frac{1}{k} & 0 & \cdots & 0 & \frac{1}{k}
  \end{pmatrix}
  \, ,
\end{equation}
while the one with respect to $a_i$ is
\begin{equation}
  \label{de_pre_jacobi_ai}
  \begin{pmatrix}
    0 & 0 & \cdots & 0 & 0
    \\
    \vdots & \vdots & \ddots & \vdots & \vdots
    \\
    - \frac{a_i^2 k^2 - 1}{\sqrt{P_J (a_i)}} & 0 & \cdots & 0 & - \frac{1}{\sqrt{P_J (a_i)}}
    \\
    \vdots & \vdots & \ddots & \vdots & \vdots
    \\
    0 & 0 & \cdots & 0 & 0
  \end{pmatrix}
  \, ,
\end{equation}
for $i = 5, \cdots, n$. We can see the non-zero elements only appear in the first and the last columns, similar to the ones in Sec.~\ref{subsec_construction}.

\subsection{The transformation to the canonical ($\varepsilon$-factorized) basis}

\label{app_can}

In this subsection, we focus on constructing the canonical basis based on the pre-canonical basis in App.~\ref{app_pre} under the maximal cut then towards the fully $\varepsilon$-factorized basis of the family. Similar to the Legendre family, we consider both of the two methods, with method~\cite{Pogel:2022vat} in App.~\ref{app_der} and constant period matrix in App.~\ref{app_period}.

\subsubsection{The construction of the canonical basis}

\label{app_der}

Similarly, we decompose the transformation to the canonical basis into two parts,
\begin{equation}
  \label{T_can_jacobi}
  {\cal T}_{\rm can}
  = {\cal T} \, {\cal T}'
  \, .
\end{equation}
The periods are 
\begin{equation}
  \label{periods_jacobi}
  \varpi_0 (k) 
  \equiv \frac{2}{\pi} \int \limits_{- 1}^{1} \frac{{\rm d} t}{\sqrt{P_J (t)}}
  = \frac{4}{\pi} K (k^2) \, ,
  \quad 
  \varpi_1 (k) 
  \equiv \frac{2 i}{\pi} \int \limits_{1}^{\frac{1}{k}} \frac{{\rm d} t}{\sqrt{P_J (t)}}
  = \frac{2}{\pi} K (1 - k^2) \, ,
\end{equation}
where $\varpi_0 (k)$ and $\varpi_1 (k)$ are holomorphic around degenerate points $k = 0$ and $k = 1$, respectively. Around degenerate point $k = 1$, the modular $\tau$ is defined as
\begin{equation}
  \tau 
  \equiv \frac{i \varpi_0 (k)}{\varpi_1 (k)} 
  = \frac{2 i K (k^2)}{K (1 - k^2)} 
  \in {\mathbb H} \, .
\end{equation}

The non-trivial elements in transformation ${\cal T}'$ are
\begin{subequations}
  \label{T_der_elements_jacobi}
  \begin{align}
    ({\cal T}')_{1, 1} 
    & = \frac{1}{\varpi_1 (k)} 
    \, ,
    \\
    ({\cal T}')_{n - 1, n - 1} 
    & = \frac{1 + \beta_{n + 1} \varepsilon}{4 \varepsilon} \varpi_1 (k)
    \, ,
    \\
    ({\cal T}')_{n - 1, j} 
    & = \frac{1}{4} \beta_{j + 3} \frac{a_{j + 3} (a_{j + 3}^2 - 1) k^2}{\sqrt{P_J (a_{j + 3})}} \varpi_1 (k)
    \, ,
    \quad 
    ({\cal T}')_{j, j}
    = 1 
    \, ,
    \quad 
    (2 \leq j \leq n - 3)
    \, ,
    \\
    ({\cal T}')_{n - 1, n - 2} 
    & = \frac{1}{4} k \varpi_1 (k) \sum_{i = 1}^{n} \beta_i a_i
    \, ,
    \quad 
    ({\cal T}')_{n - 2, n - 2}
    = 1
    \, ,
    \\
    ({\cal T}')_{n - 1, 1}
    & = - \frac{1}{4 \varepsilon} \left\{(1 - k^2) \varpi_1 (k) + k (1 - k^2) \varpi_1^\prime (k) - \varepsilon \sum_{i = 1}^{n} \beta_i [1 + k^2 (a_i^2 - 1)] \varpi_1 (k)\right\}
    \, .
  \end{align}
\end{subequations}
Under the transformation, the integral $I^{\rm pre}_{n - 1}$ is transformed to
\begin{equation}
  \label{precan_new_jacobi}
  \Big({\cal T}' \, \vec{I}^{\rm pre}\Big)_{n - 1}
  = \frac{1}{\varepsilon} \frac{\varpi_1 (k)^2}{\pi W_k} \partial_k \bigg[\frac{1}{\varpi_1 (k)} I_1\bigg] 
  \, ,
\end{equation}
where $W_k$ is the Wronskian of $\varpi_0$ and $\varpi_1$ with respect to $k$
\begin{equation}
  \label{wronskian_jacobi}
  W_k
  = \varpi_1 \partial_k \varpi_0 - \varpi_0 \partial_k \varpi_1 
  = \frac{4}{\pi k (1 - k^2)}
  \, .
\end{equation}

The non-zero elements in $\tilde{\bm A}^{(0)}_{a_i}$ are given by
\begin{subequations}
  \label{ai_de_0_new}
  \begin{align}
    (\tilde{\bm A}^{(0)}_{a_i})_{i - 3, 1}
    & = - \frac{k [k (a_i^2 - 1) \varpi_1 (k) + (1 - k^2) \varpi_1^\prime (k)]}{\sqrt{P_J (a_i)}}
    \, ,
    \\
    (\tilde{\bm A}^{(0)}_{a_i})_{n - 1, 1}
    & = - \beta_i \frac{a_i k^2 (1 - k^2) [\varpi_1 (k) + k \varpi^\prime_1 (k)] \varpi_1 (k)}{4 (a_i^2 k^2 - 1)}
    \, ,
    \\
    (\tilde{\bm A}^{(0)}_{a_i})_{n - 1, i - 3}
    & = - \beta_i \frac{k (1 - k^2) [a_i^2 k \varpi_1 (k) + (a_i^2 k^2 - 1) \varpi^\prime_1 (k)]}{4 (a_i^2 k^2 - 1) \sqrt{P_J (a_i)}}
    \, ,
  \end{align}
\end{subequations}
where $5 \leq i \leq n$. Elements $(\tilde{\bm A}^{(0)}_{a_i})_{i - 3, 1}$ can be removed with function $\vartheta_1 (a_i, k)$, and its explicit form is given by
\begin{equation}
  \label{elliptictheta1_jacobi}
  \vartheta_1 (a_i, k) 
  \equiv - \frac{2 \sqrt{P_J (a_i)}}{\pi a_i} \left[K (1 - k^2) + \frac{1}{a_i^2 - 1} \Pi \left(\frac{a_i^2 (1  - k^2)}{a_i^2 - 1} , 1 - k^2\right)\right]
  \, ,
\end{equation}
and its counterpart $\vartheta_0 (a_i, k)$ is
\begin{equation}
  \label{elliptictheta0_jacobi}
  \vartheta_0 (a_i, k) 
  \equiv - \frac{4 \sqrt{P_J (a_i)}}{\pi a_i} \Pi (1/a_i^2, k^2)
  \, .
\end{equation}

The diagonal elements in ${\cal T}$ are all equal to $1$, and other non-trivial elements are
\begin{subequations}
  \label{T_elements_jacobi}
  \begin{align}
    {\cal T}_{j, 1}
    & = - \vartheta_1 (a_{j + 3}, k)
    \, ,
    \\
    {\cal T}_{n - 1, j} 
    & = - \frac{1}{4} \beta_{j + 3} \left[\vartheta_1 (a_{j + 3}, k) + \frac{a_{j + 3} (a_{j + 3}^2 - 1) k^2 \varpi_1 (k)}{\sqrt{P_J (a_{j + 3})}}\right]
    \, ,
    \quad 
    (2 \leq j \leq n - 3)
    \, ,
    \\
    {\cal T}_{n - 1, 1}
    & = \frac{1}{8} \left\{\sum_{i = 5}^{n} \beta_i \vartheta_1 (a_i, k)^2 - \sum_{i = 1}^{n} \beta_i [1 + k^2 (a_i^2 - 1)] \varpi_1 (k)^2\right\}
    \, .
  \end{align}
\end{subequations}
With ${\cal T}'$ from Eq.~\eqref{T_der_elements_jacobi} and ${\cal T}$ from Eq.~\eqref{T_elements_jacobi}, the transformation from pre-canonical basis $\vec{I}^{\rm pre}$ to canonical basis ${\cal T}_{\rm can}$ can be obtained directly. Acting ${\cal T}_{\rm can}$ on Eq.~\eqref{pre_basis_jacobi}, we obtain a canonical basis in the Jacobi family. The canonical basis and the connection matrix here share similar properties as the ones in Legendre family.

We can decompose the total transformation ${\cal T}_{\rm can}$ into an $\varepsilon$-dependent part and an $\varepsilon$-independent part as
\begin{equation}
  \label{T_can_decomp_jacobi}
  {\cal T}_{\rm can} 
  = {\cal T}_{\rm can}^{(0)} \, {\cal T}_{\rm can}^{(\varepsilon)}
  \, ,
\end{equation}
where
\begin{equation}
  \label{T_can_eps_jacobi}
  {\cal T}_{\rm can}^{(\varepsilon)}
  = 
  \begin{pmatrix}
    \frac{1}{\varpi_1 (k)} & 0 & \cdots & 0 & 0
    \\
    0 & 1 & \cdots & 0 & 0
    \\
    \vdots & \vdots & \ddots & \vdots & \vdots
    \\
    0 & 0 & \cdots & 1 & 0
    \\
    - \frac{(1 - k^2) [\varpi_1 (k) + k \varpi_1^\prime (k)]}{4 \varepsilon} & 0 & \cdots & 0 & \frac{(1 + \beta_{n + 1} \varepsilon) \varpi_1 (k)}{4 \varepsilon}
  \end{pmatrix}
  \, .
\end{equation}

\subsubsection{The construction of an $\varepsilon$-factorized basis with constant period matrix}

\label{app_period}

With the same methods in App.~\ref{app_period_eps}, we first make the differential equations for the integrals of the third kind $\varepsilon$-factorized with transformation
\begin{equation}
  \label{T1_jacobi}
  \phi_{i - 3}^{\rm pre} 
  \to \big(\,\mathcal{T}_1 \, \vec{\phi}^{\rm pre}\,\big)_{i - 3} 
  = \phi_{i - 3}^{\rm pre} - E (\sin^{- 1} a_i , k^2) \phi_1^{\rm pre} + F (\sin^{- 1} a_i , k^2) \phi_{n - 1}^{\rm pre} 
  \, ,
\end{equation}
with $i = 5, \cdots, n$. The transformation is essentially the same as Eq.~\ref{eq:T1}, and the only difference arises in different choice of pre-canonical basis integral for the second kind, which is immaterial. Next we only need to focus on the block for $\phi_1^{\rm pre}$ and $\phi_{n - 1}^{\rm pre}$, whose period matrix is
\begin{equation}
  \label{period_pre_app}
  {\cal P}
  = 
  \begin{pmatrix}
    4 K (k^2) & - 2 i K (1 - k^2)  
    \\
    4 E (k^2) & - 2 i [K (1 - k^2) - E (1 - k^2)]
  \end{pmatrix}
  \, ,
\end{equation}
with which the transformation matrix ${\cal T}_2$ to fully $\varepsilon$-factorized basis is straightforward to construct. 

The transformation from Eq.~\eqref{pre_basis_jacobi} to the $\varepsilon$-factorized basis is
\begin{align}
  \label{T_trans_app}
  {\cal T}
  & = {\cal T}_2 \, {\cal T}_1
  \\ & = 
  \begin{pmatrix}
    E (1 - k^2) - K (1 - k^2) & 0 & \cdots & 0 & 0 & K (1 - k^2) 
    \\
    - E (\sin^{- 1} a_5, k^2) & 1 & \cdots & 0 & 0 & F (\sin^{- 1} a_5, k^2)
    \\
    \vdots & \vdots & \ddots & \vdots & \vdots & \vdots
    \\
    - E (\sin^{- 1} a_n, k^2) & 0 & \cdots & 1 & 0 & F (\sin^{- 1} a_n, k^2)
    \\
    0 & 0 & \cdots & 0 & 1 & 0
    \\
    E (k^2) & 0 & \cdots & 0 & 0 & - K (k^2)
  \end{pmatrix} 
    \, .
\end{align}
Similarly, with this transformation, the period matrix is a constant matrix.

\subsubsection{From Jacobi family back to the original family}

\label{app_jacobi2original}

We still need to find the transformation which transforms the original family to the Jacobi family. The transformation can be a M\"obius transformation or a quadratic transformation. Now let's come to the former first.

\subsubsection*{M\"obius transformation}

\label{app_M\"obius}

Basically, we can combine two M\"obius transformations where one of them is exactly the same as the one in Sec.~\ref{subsec_legendre2original} and the other is its inverse and treat the Jacobi family as the original family. To be explicit, the M\"obius transformation can be expressed as
\begin{equation}
  z 
  \mapsto 
  x 
  = T_{4L} (z)
  \mapsto 
  t 
  = T_{LJ} (x)
  = T_{LJ} \circ T_{4L} (z) 
  = S (z)
  \, .
\end{equation}
We've known the transformation from the original family to the Jacobi family is 
\begin{equation}
  \label{T_42L}
  x
  = T_{4L} (z)
  = \frac{(z - c_2) c_{14}}{(z - c_1) c_{24}}
  \, ,
\end{equation}
and specially, if we specify the original family as the Jacobi family, then it becomes
\begin{equation}
  \label{T_J2L}
  x
  = T_{JL} (t)
  = \frac{(t - a_2) a_{14}}{(t - a_1) a_{24}}
  = \frac{(t + 1) \frac{2}{k}}{(t + \frac{1}{k}) \frac{1 + k}{k}}
  = \frac{2 k (1 + t)}{(1 + k) (1 + k t)}
  \, .
\end{equation}
While $T_{LJ}$ is nothing but the inverse of $T_{JL}$, so we can find
\begin{equation}
  \label{T_L2J}
  t
  = T_{LJ} (x)
  = \frac{(1 + k) x - 2 k}{k [2 - (1 + k) x]}
  \, ,
\end{equation}
and we combine it with $T_{4L}$ to get $S$. However, we can see we will obtain a complicated transformation even without expressing $k$ with $c_i$'s, which may make the expression worse. Thus, even we can construct the pre-canonical basis in the Jacobi family, the corresponding basis in the original family seems not to be easy to use, due to the complexity of the M\"obius transformation $S (z)$.

We consider the M\"obius transformation from a generic family to the Jacobi family above, which seems to be complicated in general. One may wonder although the transformation is cumbersome for generic cases, are there some special configurations of $c_i$'s such that the transformation is a simple linear one. Fortunately, the answer is positive. For example, suppose we have the four branch points (the ones related to the shape of the elliptic curve, instead of the ones corresponding to marked points) in a symmetric manner
\begin{equation}
  \label{symmetric}
  c_{12} 
  = c_{34}
  \, .
\end{equation}
The M\"obius transformation to its Jacobi family is easy to construct since it is just a combination of translation and dilation. To be more explicit, a possible M\"obius transformation is
\begin{equation}
  \label{S_symmetric}
  z 
  \mapsto t 
  = S (z)
  = \frac{2 z - (c_2 + c_3)}{c_{32}}
  \, .
\end{equation}

With transformation $S (z)$, and we apply the symmetric relations of the branch points, the branch points are mapped to 
\begin{equation}
  \label{symmetric_branch_points}
  a_1 
  = - \frac{c_{14}}{c_{23}}
  = - \frac{1}{k}
  \, ,
  \quad 
  a_2 
  = - 1
  \, ,
  \quad 
  a_3 
  = + 1
  \, ,
  \quad 
  a_4
  = + \frac{c_{14}}{c_{23}}
  = + \frac{1}{k}
  \, ,
  \quad 
  a_i 
  = \frac{c_{2i} + c_{3i}}{c_{23}}
  \, ,
\end{equation}
where $i \geq 5$ and $c_i$'s correspond to the marked points, and a special case is $a_{\infty} = \infty$. This behavior is as expected since in the transformation $S (z)$ Eq.~\eqref{S_symmetric} there is no inversion which is the only transformation that alters the point at the infinity. We can easily read out from Eq.~\eqref{symmetric_branch_points} that the corresponding $k$ is
\begin{equation}
  \label{k_symmetric}
  k 
  = \frac{c_{23}}{c_{14}}
  \, .
\end{equation}
Besides, we also have relations
\begin{equation}
  \label{z2t_app}
  {\rm d} z
  = \frac{c_{32}}{2} {\rm d} t
  \, , 
  \quad
  z - c_i
  = \frac{c_{32}}{2} (t - a_i) 
  \, ,
\end{equation}
where $i = 1, \cdots \, , n$.
Plugging the relations into $u (z)$, we find
\begin{equation}
  u (z) \, {\rm d} z
  \propto
  u (t) \, {\rm d} t
  \, .
\end{equation}

In order to express the pre-canonical basis in the original family, we also need the relations for the inverse transformation, which can be easily read out from Eq.~\eqref{z2t_app}.
The pre-canonical basis in the original family is then straightforward
\begin{subequations}
  \label{pre_basis_app}
  \begin{align}
    \phi_1^{\rm pre} 
    & = \frac{c_{41}}{2} \frac{{\rm d} z}{\sqrt{P_4 (z)}}
    \, ,
    \\
    \phi_{i - 3}^{\rm pre}
    & = \frac{1}{z - c_i} \frac{\sqrt{P_4 (c_i)}}{\sqrt{P_4 (z)}} {\rm d} z 
    \, ,
    \\
    \phi_{n - 2}^{\rm pre}
    & = \left(z - \frac{c_2 + c_3}{2}\right) \frac{{\rm d} z}{\sqrt{P_4 (z)}}
    \, ,
    \\
    \phi_{n - 1}^{\rm pre}
    & = \frac{2}{c_{14}} (z - c_1) (z - c_4) \frac{{\rm d} z}{\sqrt{P_4 (z)}} 
    \, ,
  \end{align}
\end{subequations}
where $i = 5, \cdots, n$.

We want to emphasize that the example mentioned above represents a very special case, where the point at infinity in the Jacobi normal form corresponds directly to the point at infinity in the original form. However, in more general cases, the point at infinity in the normal form may differ from the point at infinity in the original form. This distinction complicates the transformation process, but it can be circumvented by employing an alternative representation for the elliptic integral of the second kind, e.g.,
\begin{equation}
  \label{second_basis}
  \phi_{n - 1}^{\rm pre}
  = \frac{1 - k^2}{k} \frac{1}{t + \frac{1}{k}} \frac{{\rm d} t}{\sqrt{P_J (t)}}
  \, ,
\end{equation}
and other differentials with a double pole at $t = - 1, + 1, + 1 / k$ are also possible. While for the ${\rm d} \log$-form, we replace it with the elliptic integral of the third kind 
\begin{equation}
  \phi_{n - 2}^{\rm pre}
  = \frac{1}{t - a_\infty} \frac{\sqrt{P_J (a_\infty)}}{\sqrt{P_J (t)}} {\rm d} t 
  \, ,
\end{equation}
where $a_\infty$ is the point which corresponds to the point at infinity $c_\infty = \infty$ in the original family.

\subsection*{Quadratic transformation}

An alternative transformation is to use a quadratic transformation instead of another M\"obius transformation, from the Legendre form in Sec.~\ref{sec_construction}. The quadratic transformation is simple to construct, the explicit form is $x \mapsto t = \sqrt{x / \lambda}$, $e_i \mapsto a_i = \sqrt{e_i / \lambda}$. Then the overall transformation from the original elliptic curve to the Jacobi form is
\begin{equation}
  \label{S_quadratic}
  z 
  \mapsto t 
  = S (z)
  = \sqrt{\frac{(z - c_2) c_{13}}{(z - c_1) c_{23}}}
  \, .
\end{equation}

With the quadratic transformation, the $\widetilde{u} (t)$ is 
\begin{equation}
  \label{quadjacobi}
  \widetilde{u} (t) 
  \propto P_{4J} (t)^{- 1 / 2} \prod_{i = 2}^{n + 1} (t^2 - a_i^2)^{- \beta_i \varepsilon}
  \, ,
\end{equation}
where the $k$ specifying the Jacobi curve here is the modulus $k^2 = \lambda$. We can see the twist of the Jacobi family is different from Eq.~\eqref{twist_jacobi}, in that the marked points $\pm a_i$ now appear in pairs. As a result, the number of the basis integrals is twice as many as the original family. To deal with that, besides pre-canonical basis integrals in Eq.~\eqref{pre_basis_jacobi}, we can construct additional ones as ${\rm d} \log$-forms. These additional integrals disentangle with those in Eq.~\eqref{pre_basis_jacobi} in the $\varepsilon^0$-part of the differential equations. These additional ${\rm d} \log$-forms do not belong to the original family, and are intrinsic in the Jacobi family~\eqref{quadjacobi}. To see that, it is easy to show that when inverse-transforming from $t$ back to the variable $z$, these ${\rm d} \log$-forms necessarily involve square-roots of the form \eqref{S_quadratic}, which do not appear in the twist of the original family. In contrast, the integrals in Eq.~\eqref{pre_basis_jacobi} would not have such square-roots under the inverse transformation, so they can indeed serve as pre-canonical basis integrals in the original family.

We want to point out that the construction in Jacobi normal form is in fact equivalent to the one in Legendre normal form in Sec.~\ref{sec_construction}, up to a constant rotation, which is as expected since the factor between ${\rm d} x / \sqrt{x (x - \lambda) (x - 1)}$ and ${\rm d} t / \sqrt{(1 - t^2) (1 - k^2 t^2)}$ is a trivial constant. 

\subsection{Application to non-planar double box family}

\label{app_npdb}

Although the simple pre-canonical basis integrals in Eq.~\eqref{pre_basis_app} ask a special original family, it turns out our first non-planar double box family in Sec.~\ref{subsec_npdb} satisfies the condition. It is important to note that the underlying elliptic curve possesses a unique property: the branch points exhibit an additional symmetry. This symmetry makes the construction in Jacobi form simpler to use.

For the branch points of the single mass case in Sec.~\ref{subsec_npdb}, we can refer back to Eq.~\eqref{branch_point_npdb}. With the M\"obius transformation in Eq.~\eqref{S_symmetric}, the variables $k$ and $a_5$ for the elliptic curve in the Jacobi normal form are
\begin{equation}
  \label{ka_app}
  k
  = \sqrt{1 - 16 y_s} 
  \, ,
  \quad
  a_5 
  = \frac{2 y_t - 1}{\sqrt{1 - 16 y_s}}
  \, .
\end{equation}
Plugging Eq.~\eqref{ka_app} into Eq.~\eqref{pre_basis_jacobi}, one can check this pre-canonical basis enjoys an arguably simpler form than the one obtained in Sec.~\ref{subsec_npdb}, and here algebraic factors only appear in $\phi_2^{\rm pre}$. We also note that the pre-canonical basis integrals in~\cite{Becchetti:2025rrz} are very similar to the ones here.

Since the explicit forms of $k$, $a_5$ are so simple that we can easily express them with kinematic variables $y_s$, $y_t$ and so is the inverse, we can apply a base transformation to take $k$ and $a_5$ as the variables of the integrals. With such base transformation, the geometric information is more explicit and we have more concise representations for the integrals and the differential equations.

In order to reach a canonical basis for the whole family, we only need to apply essentially the same strategy in Sec.~\ref{subsec_npdb}. One can check the two constructions are equivalent. However, the involved pre-canonical basis and ${\cal T}_{\rm can}$ have simpler arguments than the ones for the basis in Sec.~\ref{subsec_npdb}. The elliptic integrals with distinct arguments can be related by non-trivial identities like the ones in App.~\ref{app_special}.

\section{The pre-canonical basis in the explicit form}

\label{app_basis}

In this appendix, we will provide the explicit form of the pre-canonical basis of Sec.~\ref{subsec_npdb} for which the differential equation is $\varepsilon$-factorized, except in the top-sector. To achieve a fully $\varepsilon$-factorized differential equation, one can apply the transformation from \eqref{T_can} within the top-sector. 

However, the terms arising from the non-elliptic sub-sectors for  $I_2$ and $I_4$ are too numerous to present in full. Therefore, we will limit the discussion to the contributions from the elliptic sectors for integrals in elliptic sectors.

\begingroup
  \allowdisplaybreaks
  \begin{align}
    I_1 & 
    = - \frac{1}{2} \varepsilon^4 \left(\sqrt{1-16 y_s} + 1\right) (I_{111110100} - I_{11111110(-1)})
    \, ,
    \\ 
    I_2 & 
    = \frac{\varepsilon^4 \sqrt{y_t \left(y_t - 1\right) \left[4 y_s+\left(y_t - 1\right) y_t\right]}}{y_t}
    \left[y_t I_{111111100} - I_{111110100} + I_{11111110(-1)}\right]
    \, ,
    \\
    I_3 & 
    = \frac{1}{4} \varepsilon^3 \left[2 y_s \left(I_{11121110(-1)} - I_{21111110(-1)}\right) 
    - 2 \varepsilon \left(y_t-1\right) I_{111110100} \right.
    \nonumber \\ & 
    \left. + 2 \varepsilon \left(y_t - 1\right) y_t I_{111111100} 
    + \varepsilon \left(2 y_t - 3\right) I_{11111110(-1)}\right]
    \, ,
    \\
    I_4 &
    = - \frac{1}{8 y_s} \varepsilon^4 \left[\left(\varepsilon \left(\sqrt{1-16 y_s}-1\right)+8y_s \left(4 \varepsilon  \left(\sqrt{1-16y_s}-1\right)+\sqrt{1-16 y_s}\right)\right) I_{111110100} \right.
    \nonumber \\ &
    \left. - 4 y_s \left(2 \left(4 \varepsilon  \left(\sqrt{1-16y_s}-1\right)+\sqrt{1-16 y_s}\right) I_{11111110(-1)} \right. \right.
    \nonumber \\ &
    \left. \left. + \left(\sqrt{1-16y_s}-1\right) \left((16y_s-1) I_{111210100} + (1-8 y_s) I_{11121110(-1)} - 8 y_s I_{21111110(-1)}
    \right. \right. \right.
    \nonumber \\ &
    \left. \left. \left. + 4 \varepsilon  (y_t-1) I_{111111100}\right)\right)\right] 
    \, ,
    \\
    M_5 & 
    = \frac{\varepsilon^4 \pi}{4 K (16 y_s)} I_{111110100}
    \, ,
    \\
    M_6 & 
    = \frac{\varepsilon^3}{8 \pi} \left[\left(2 \left(E (16 y_s) + \left(48 y_s \varepsilon - 2 \varepsilon +16 y_s-1\right) K (16 y_s)\right) - \pi^2 \varepsilon\right) I_{111110100}\right.
    \nonumber \\ &
    \left. + 16 \left(1 - 16 y_s\right) y_s K (16 y_s) I_{111210100}\right]
    \, ,
    \\
    M_7 & 
    = \frac{1}{64} \varepsilon^3 \left(4 y_s y_t I_{111201100} 
    - \left(y_t - 1\right) I_{201101100} 
    + y_t \left(y_t - 1\right) I_{111102000}\right.
    \nonumber \\ &
    \left. - y_t I_{110201100}\right)
    \, ,
    \\
    M_8 & 
    = - \frac{1}{8} \varepsilon^4 \sqrt{y_s \left(y_t - 1\right) y_t} 
    I_{111101100} 
    \, ,
    \\ 
    M_9 & 
    = - \frac{1}{16} \varepsilon^4 y_t I_{110111100} 
    \, ,
    \\
    M_{10} & 
    = - \frac{1}{16} \varepsilon^4 \left(y_t - 1\right)
    I_{111011100} 
    \, ,
    \\
    M_{11} &
    = - \frac{1}{16} \varepsilon^3 \sqrt{y_t (y_t  - 4 y_s)} 
    I_{010211100} 
    \, ,
    \\
    M_{12} & 
    = \frac{1}{4} \varepsilon^2 \left[\varepsilon \left(2 \varepsilon - 1\right) I_{010111100} 
    + \varepsilon \left(y_t - 4 y_s\right) I_{010211100}
    - I_{002020100}\right]
    \, ,
    \\
    M_{13} & 
    = \frac{1}{16} \varepsilon^3 \sqrt{\left(y_t - 1\right) (4 y_s + y_t - 1)}
    I_{201011100} 
    \, ,
    \\
    M_{14} & 
    = \frac{1}{8} \varepsilon^2
    \left[I_{202001000} 
    - 2 \varepsilon \left(I_{102010100} + I_{20101110(-1)}\right)\right]
    \, ,
    \\
    M_{15} & 
    = - \frac{1}{16} \varepsilon^3 \sqrt{y_t [4 y_s \left(y_t - 1\right) + y_t]} I_{110201100} 
    \, ,
    \\
    M_{16} &  
    = \frac{1}{4} \varepsilon^4 \left(y_t - 1\right) I_{110101100} 
    \, , 
    \\
    M_{17} & 
    = - \frac{1}{16} \varepsilon^3 \sqrt{\left(y_t - 1\right) (4 y_s y_t + y_t - 1)} I_{201101100}
    \, ,
    \\
    M_{18} & 
    = \frac{1}{4} \varepsilon^4 y_t I_{101101100} 
    \, ,
    \\
    M_{19} & 
    = \frac{\varepsilon^3}{8 (\varepsilon - 1) \left(y_t - 1\right)} 
    \left[\left(2 y_s + (3 \varepsilon - 1) y_t\right) I_{020201000}\right.
    \nonumber \\ &
    \left. - \left(2 y_s - 3 \varepsilon \left(y_t - 1\right) + y_t - 1\right)
    I_{202001000}
    + \varepsilon \left(8 y_s - 2 y_t\right) I_{010202000}\right.
    \nonumber \\ &
    \left. + 2 \left((1 - \varepsilon) y_s I_{121101000} 
    - \varepsilon \left(4 y_s + y_t - 1\right) I_{201002000}\right.\right.
    \nonumber \\ &
    \left.\left. + \varepsilon \left((2 \varepsilon - 1) y_s + (\varepsilon - 1) y_t\right) I_{110201000}\right.\right.
    \nonumber \\ &
    \left.\left. + \varepsilon \left(y_s - y_t + \varepsilon \left(- 2 y_s + y_t - 1\right) + 1\right) I_{112001000}\right)\right]
    \, ,
    \\
    M_{20} & 
    = \frac{1}{16} \varepsilon^3 \sqrt{\left(y_t-1\right) y_t [4 y_s + \left(y_t - 1\right) y_t]} I_{111102000}
    \, ,
    \\
    M_{21} & 
    = - \frac{1}{4} \varepsilon^4 I_{111101000}
    \, ,
    \\
    M_{22} & 
    = - \frac{1}{4} \varepsilon^4 I_{111010100}
    \, ,
    \\
    M_{23} & 
    = - \frac{1}{4} \varepsilon^4 I_{111100100}
    \, ,
    \\
    M_{24} & 
    = \frac{1}{16} \varepsilon^2 \sqrt{1 - 4 y_s} \left[\frac{\varepsilon}{2 y_s (1 + 2 \varepsilon)} I_{022000000} + I_{102010200}\right]
    \, ,
    \\
    M_{25} & 
    = - \frac{1}{4} \varepsilon^3 I_{102010100}
    \, ,
    \\
    M_{26} & 
    = \frac{1}{4} \varepsilon^3 y_t I_{110201000}
    \, ,
    \\
    M_{27} & 
    = - \frac{1}{4} \varepsilon^3 \left(y_t - 1\right) I_{112001000} 
    \, ,
    \\
    M_{28} & 
    = - \frac{1}{4} \varepsilon^3 I_{110200100}
    \, ,
    \\
    M_{29} & 
    =\frac{1}{4} \varepsilon^2 \sqrt{y_t (y_t - 4 y_s)}
    \left(2 I_{010202000} + I_{020201000}\right)
    \, ,
    \\
    M_{30} & 
    = \frac{1}{4} \varepsilon^2 y_t I_{020201000}
    \, ,
    \\
    M_{31} & 
    = - \frac{1}{4} \varepsilon^2 \sqrt{\left(y_t - 1\right) (4 y_s + y_t - 1)} \left(2 I_{201002000} + I_{202001000}\right)
    \, ,
    \\
    M_{32} & 
    = - \frac{1}{4} \varepsilon^2 \left(y_t - 1\right) I_{202001000}
    \, ,
    \\
    M_{33} & 
    = - \frac{1}{4} \varepsilon^2 \sqrt{4 y_s + 1} \left(2 I_{200100200} + I_{200200100}\right)
    \, ,
    \\
    M_{34} & 
    = - \frac{1}{4} \varepsilon^2 I_{200200100}
    \, ,
    \\
    M_{35} & 
    = - \frac{1}{4} \varepsilon^2 I_{002020100}
    \, ,
    \\
    M_{36} & 
    = \varepsilon^2 I_{022000000}
    \, .
  \end{align}
\endgroup

\section{Special relations of elliptic integrals}

\label{app_special}

In this appendix, we will briefly introduce some special relations that changes the arguments of elliptic integrals. We have changed the arguments $\lambda =4 k / (1 + k)^2$ to $k^2$ from Eq.~\eqref{special_relation_npdb} to Eq.~\eqref{special_relation_sim_npdb}, and we know the two kinds of arguments are the ones for Legendre family in Sec.~\ref{sec_construction} and Jacobi family in App.~\ref{app_jacobi}, respectively. A natural way to relate the elliptic integrals with these two kinds of arguments is to consider the M\"obius transformation Eq.~\eqref{T_L2J} which bridges them, combined with the quadratic transformation in Eq.~\eqref{Legendre2Jacobi}. 

With Eqs.~\eqref{T_L2J} and~\eqref{Legendre2Jacobi}, we obtain a quadratic transformation which transforms a standard elliptic integral with parameter $k^2$ to the one with parameter $\lambda = 4 k / (1 + k)^2$. The transformation is given by
\begin{equation}
  \label{J2J_quadratic}
  \tilde{t}^2
  = \frac{x}{\lambda}
  = \frac{(1 + k) (1 + t)}{2 (1 + k t)}
  \, .
\end{equation}
Under this transformation, we have relations
\begin{equation}
  \label{J2L_differential}
  \frac{{\rm d} t}{\sqrt{(1 - t^2) (1 - k^2 t^2)}}
  = \frac{2}{1 + k} \frac{{\rm d} \tilde{t}}{\sqrt{(1 - \tilde{t}^2) (1 - \lambda \tilde{t}^2)}}
  \, ,
\end{equation}
and 
\begin{equation}
  \label{J2L_mp}
  \tilde{a}^2
  = \frac{e}{\lambda}
  = \frac{(1 + k) (1 + a)}{2 (1 + k a)}
  \, .
\end{equation}
We will mainly focus on the region $0 < k < 1 < 1 / k < a$ for simplicity, though the relations we obtain hold for a larger domain.

We can relate complete elliptic integrals of the first kind with arguments $k^2$ and $\lambda$
\begin{equation}
  \label{K_relation}
  K (k^2)
  = \frac{1}{1 + k} K (\lambda)
  \, ,
  \quad 
  K (1 - k^2)
  = \frac{2}{1 + k} K (1 - \lambda)
  \, .
\end{equation}
The inverse identities are
\begin{equation}
  \label{K_inverse}
  K (\lambda)
  = (1 + k) K (k^2)
  \, ,
  \quad 
  K (1 - \lambda)
  = \frac{1 + k}{2} K (1 - k^2)
  \, .
\end{equation}
We can also use the periods defined in Eqs.~\eqref{periods} and~\eqref{periods_jacobi} to rewrite the relations above, which are
\begin{equation}
  \label{K_period}
  \varpi_0 (\lambda)
  = (1 + k) \varpi_0 (k)
  \, ,
  \quad
  \varpi_1 (\lambda)
  = (1 + k) \varpi_1 (k)
  \, .
\end{equation}
We can simply apply the derivation to the both sides of the identities above to obtain the relations for complete elliptic integrals of the second kind, which is rather straightforward, so we would not bother to write them down explicitly here.

Similarly, we can also derive the relations for complete elliptic integrals of the third kind, which are
\begin{equation}
  \label{Pi_period}
  \vartheta_0 (e, \lambda)
  = \vartheta_0 (a, k) + \frac{k \sqrt{P_J (a)} \varpi_0 (k)}{1 + a k}
  \, ,
  \quad
  \vartheta_1 (e, \lambda)
  = \vartheta_1 (a, k) + \frac{k \sqrt{P_J (a)} \varpi_1 (k)}{1 + a k} - 1
  \, .
\end{equation}
For the relations expressed with complete elliptic integrals of the third kind explicitly, we just need to plug the definitions explicitly.

The identities above establish relationships between complete elliptic integrals with arguments $(k^2, 1 - k^2)$ and $(\lambda, 1 - \lambda)$. A natural question arises: can we find other similar relations that connect different sets of arguments? The answer is affirmative. Recall that the identities above were derived using the M\"obius transformation in Eq.~\eqref{T_J2L} and its inverse Eq.~\eqref{T_L2J}. However, these are not the only M\"obius transformations that map the Jacobi curve to the Legendre curve. By reordering the $c_i$'s in Eqs.~\eqref{T_J2L} and~\eqref{T_L2J} , we can identify additional M\"obius transformations, which in turn allow us to derive other similar relations. The resulting arguments on the RHS are
\begin{equation}
  \label{6_cross_ratios}
  \left(\lambda, 1 - \lambda\right), \left(\frac{1}{\lambda}, 1 - \frac{1}{\lambda}\right), \left(\frac{1}{1 - \lambda}, 1 - \frac{1}{1 - \lambda}\right) 
  \, .
\end{equation}
These arguments correspond to the six cross ratios for a single elliptic curve. Moreover, since the LHS of the relations are fixed, such as $K (k^2)$ and $K (1 - k^2)$, we can also derive relations for elliptic integrals whose arguments are among those in Eq.~\eqref{6_cross_ratios}. These relations can be interpreted as reflecting different choices of the modulus $\lambda$ for a single elliptic curve. This interpretation can be verified by examining the \textit{Klein's j-invariants}.

In Eq.~\eqref{K_relation}, there are two more different arguments $k^2$ and $1 - k^2$ on the LHS which do not belong to Eq.~\eqref{6_cross_ratios}, implying that these relations connect two distinct elliptic curves. However, there exist specific relationships between these two elliptic curves. As shown in App.~\ref{app_jacobi2original}, the two elliptic curves are connected via a quadratic transformation, up to a M\"obius transformation, which does not alter the elliptic curve. Consequently, the two elliptic curves are isogenic, as evidenced by
\begin{equation}
  \tau_M 
  = \frac{i K (16 y_s)}{2 K (1 - 16 y_s)} 
  = 2 \frac{i K \left(1 - \frac{4 \sqrt{1 - 16 y_s}}{\left(1 + \sqrt{1 - 16 y_s}\right)^2}\right)}{2 K \left(\frac{4 \sqrt{1 - 16 y_s}}{\left(1 + \sqrt{1 - 16 y_s}\right)^2}\right)} 
  = 2 \tau_Q
  \, ,
\end{equation}
where we've applied Eq.~\eqref{K_relation} with arguments in Eqs.~\eqref{ka_app} and~\eqref{variable_npdb} for a concrete example.

\end{appendix}

% references
{\footnotesize
\bibliography{refs}
\bibliographystyle{JHEP.bst}
}

\end{document}